\documentclass[letterpaper,10pt]{article}

\usepackage{jheppub}

\usepackage{physics}

\usepackage[export]{adjustbox}

\def\be{\begin{equation}}
\def\ee{\end{equation}}
\def\ba{\begin{aligned}}
\def\ea{\end{aligned}}
\def\p{\partial}
\def\oh{\mathcal{O}}
\def\ah{\mathcal{A}}
\def\nh{\mathcal{N}}

\title{Large deformations of Tr($\Phi^3$) and the world at infinity}

\abstract{The amplitudes of the non-linear sigma model can be obtained from those of Tr($\Phi^3$) theory by sending the kinematic (Mandelstam) variables to infinity in a certain direction. In this paper we characterize the behavior of Tr($\Phi^3$) amplitudes under a general class of large kinematic shifts called $g$-vector shifts. The objects that live in this world at infinity retain certain key amplitude-like properties, most notably factorization, and admit descriptions in terms of polytopes, but they are not generally amplitudes of any cognizable theory. We identify particular $g$-vector shifts that lead at infinity to mixed amplitudes involving two pions and any number of scalars, allowing us to provide polytopal descriptions of these amplitudes.}

\author[a]{Shruti Paranjape,}
\emailAdd{shruti\_paranjape@brown.edu}

\author[a]{Marcos Skowronek,}
\emailAdd{marcos\_skowronek\_santos@brown.edu}

\author[a,b]{Marcus Spradlin,}
\emailAdd{marcus\_spradlin@brown.edu}

\author[a]{Anastasia Volovich}
\emailAdd{anastasia\_volovich@brown.edu}

\affiliation[a]{Department of Physics,
    Brown University,
    Providence,
    RI 02912,
    USA
}

\affiliation[b]{Brown Theoretical Physics Center,
    Brown University,
    Providence,
    RI 02912,
    USA
}

\begin{document}
	
	\maketitle
	
	\section{Introduction}\label{sec: introduction}
	
	Different descriptions of scattering amplitudes have proved useful in illuminating many of their interesting properties. The geometric description of amplitudes, via the volume or canonical form on a positive geometry (like a polytope) can bypass the traditional approach of summing over a combinatorially large number of Feynman diagrams. Such constructions have found numerous applications to amplitudes in a wide range of theories including $\mathcal{N}=4$ supersymemtric Yang-Mills (SYM) theory \cite{Arkani-Hamed:2013jha,Damgaard:2019ztj}, 3d ABJM theory \cite{He:2023rou}, Tr($\Phi^3$) theory \cite{Arkani-Hamed:2017mur}, correlators in cosmology \cite{Arkani-Hamed:2024jbp} and in $\mathcal{N}=4$ SYM \cite{Eden:2017fow}, as well as for Wilson coefficients of EFTs \cite{Arkani-Hamed:2020blm}.
	
	Recently, a new construction of amplitudes in Tr($\Phi^3$) theory via \emph{kinematic surfaces} \cite{Arkani-Hamed:2019vag,Arkani-Hamed:2023lbd,Arkani-Hamed:2023mvg,Arkani-Hamed:2023swr,Arkani-Hamed:2023jry,Arkani-Hamed:2024nhp,Arkani-Hamed:2024vna,Arkani-Hamed:2024yvu,Arkani-Hamed:2024fyd,De:2024wsy,Arkani-Hamed:2024tzl,Arkani-Hamed:2024pzc} has revealed its connection to the $SU(N)$ non-linear sigma model (NLSM) of pions \cite{Arkani-Hamed:2024nhp, Arkani-Hamed:2024yvu,Dong:2024klq}, as well as to gluons in Yang-Mills (YM) theory \cite{Arkani-Hamed:2023jry}. The former involves a large deformation of the kinematic variables, referred to as the $\delta$-shift. The $\delta$-shift selects two subsets of kinematic variables to shift to positive or negative infinity. Amplitudes in Tr($\Phi^3$) theory only fall off at infinity in special directions. Thus the $\delta$-shift is surprising in two ways: it is well-behaved despite being an infinitely large deformation of Tr($\Phi^3$) theory, and it connects theories with different numbers of derivatives and valences\footnote{Here valence refers to the number of fields that interact in the vertices of the theory i.e.~3 for Tr$(\Phi^3)$ and all even numbers greater than 2 for the NLSM.}.
	
	The behavior of amplitudes at infinity in kinematic space are of great interest in a variety of theories. The suppression of gluon and graviton tree amplitudes under a large BCFW shift is what allows them to be recursively constructed \cite{Britto:2004ap, Britto:2005fq, Benincasa:2007qj, Arkani-Hamed:2008bsc}. The enhancement of such behavior has been studied in the context of loop integrands in supergravity \cite{Bern:2012gh, Bern:2014sna, Bern:2017lpv, Herrmann:2018dja, Edison:2019ovj} and SYM \cite{Bourjaily:2018omh, Brown:2022wqr}, and explains the existence of bonus relations in these theories \cite{Bedford:2005yy, Cachazo:2005ca}. Good behavior of amplitudes or integrands in the ultraviolet (UV) kinematic region is also linked to the existence of symmetries, e.g. dual superconformal symmetry in SYM theory is a consequence of the absence of poles at infinity \cite{Drummond:2008vq}. In cases where amplitudes scale badly, such as in EFTs, it is possible to systematically introduce subtractions in order to eliminate residues at infinity \cite{Cheung:2014dqa, Cheung:2015ota}. Some large deformations of positive geometries such as the associahedron have been shown to be well-behaved \cite{Aneesh:2019ddi, Aneesh:2019cvt, Jagadale:2021iab}. The behavior of amplitudes under large deformations has also been shown to be linked to the presence of hidden zeros \cite{Rodina:2024yfc} and near-zero splitting \cite{inprogress:Jones}. Nonetheless, all of these studies of the world at infinity necessarily involve special directions.
	
	In this paper, we introduce and study a class of large kinematic deformations of Tr($\Phi^3$) that we call $g$-vector shifts, which generalize the $\delta$-shift that gives pion amplitudes in the NLSM. In particular cases, these deformations also coincide with the ones used in \cite{He:2018svj,Yang:2019esm} to derive recursion relations in $\Phi^3$ theory. These shifts naturally arise in the context of the associahedron and surfacehedron, polytopes that give the amplitudes of Tr$(\Phi^3)$ theory at tree- and loop-level respectively. We provide a full characterization of the functions at infinity that result from these shifts. We show that the resulting objects living at kinematic infinity retain some important amplitude-like properties; in particular they still factorize correctly on their remaining poles. This leads to the natural question of when these functions at infinity are actual amplitudes.
	
	Indeed a fundamental idea behind the amplitude bootstrap program is that functions of kinematic variables with the correct analytic properties (our result at infinity being an example) correspond to some Lagrangian. A special case is when the function at infinity coincides with amplitudes in a theory of pions coupled to Tr($\Phi^3$) scalars. These are the same amplitudes that were first observed in \cite{Cachazo:2016njl} as coefficients of the soft expansion of the NLSM; the Lagrangian of this theory is given in \cite{Mizera:2018jbh}. In the further specialized case of an amplitude of two pions and any number of scalars, there is a remarkable simplification that occurs. Underlying this simplification is the idea that large deformations can be seen not just at the functional level, but also as the shifting of facets (codimension-1 boundaries) of the associahedron to infinity. Indeed, deformations resulting in this class of amplitudes give the combinatoric structure of one lower-point or a product of two lower-point associahedra. Thus, we can construct a polytope description of these two-pion amplitudes. In all the examples of geometric descriptions discussed above, we note that none of the theories are EFTs i.e. none have higher-derivative operators. Our example breaks this pattern, giving an example of a polytopal description of a theory with a 2-derivative 4-scalar interaction.
	
	In Section \ref{sec: review kin space}, we review the construction of variables from kinematic meshes and surfaces. In Section \ref{sec:AatInfinity}, we define $g$-vector shifts and present our main result -- a full characterization of the result one obtains by taking the amplitudes of Tr($\Phi^3$) to infinity along such shifts. We also discuss the pion amplitudes that result from some large deformations. In Section \ref{sec:associahedron interpretation}, we discuss how to understand our large kinematic deformations geometrically, via the associahedron. In Section \ref{sec:polytope}, we present a new polytope description for amplitudes involving two pions and an arbitrary number of scalars. Finally, we discuss open problems for future work in Section \ref{sec:discussion}.

	\section{Review of kinematic space}\label{sec: review kin space}
	
	In this section we review following \cite{Arkani-Hamed:2019vag} the kinematic mesh picture that describes the kinematics of amplitudes in $\Tr(\Phi^3)$ theory and in particular the so-called $g$-vectors. These will allow us to set up the particular deformations of Tr($\Phi^3$) amplitudes that are the primary focus of this work.
	
	We consider tree-level amplitudes in the theory of colored massless scalar fields transforming in the adjoint representation of $U(N)$, i.e., an $N \times N$ matrix valued field $\Phi$, with cubic interactions described by the Lagrangian:
	\be \mathcal{L} = \Tr(\p_\mu \Phi \p^\mu \Phi) + g\Tr(\Phi^3).\ee
	The $n$-point amplitude in Tr($\Phi^3$) theory is a function of Lorentz-invariant scalar products $p_i \cdot p_j$ of the $n$ momenta $p_i$. The space spanned by these Mandelstam variables is called the kinematic space $\mathcal{K}_n$. Assuming that the dimension of spacetime is sufficiently large ($>n{-}2$), momentum conservation and the on-shell conditions $p_i^2 = 0$ imply that $\dim(\mathcal{K}_n)=n(n{-}3)/2$.
	
\begin{figure}[ht]
		\centering
		\includegraphics[width=0.23\textwidth,valign=c]{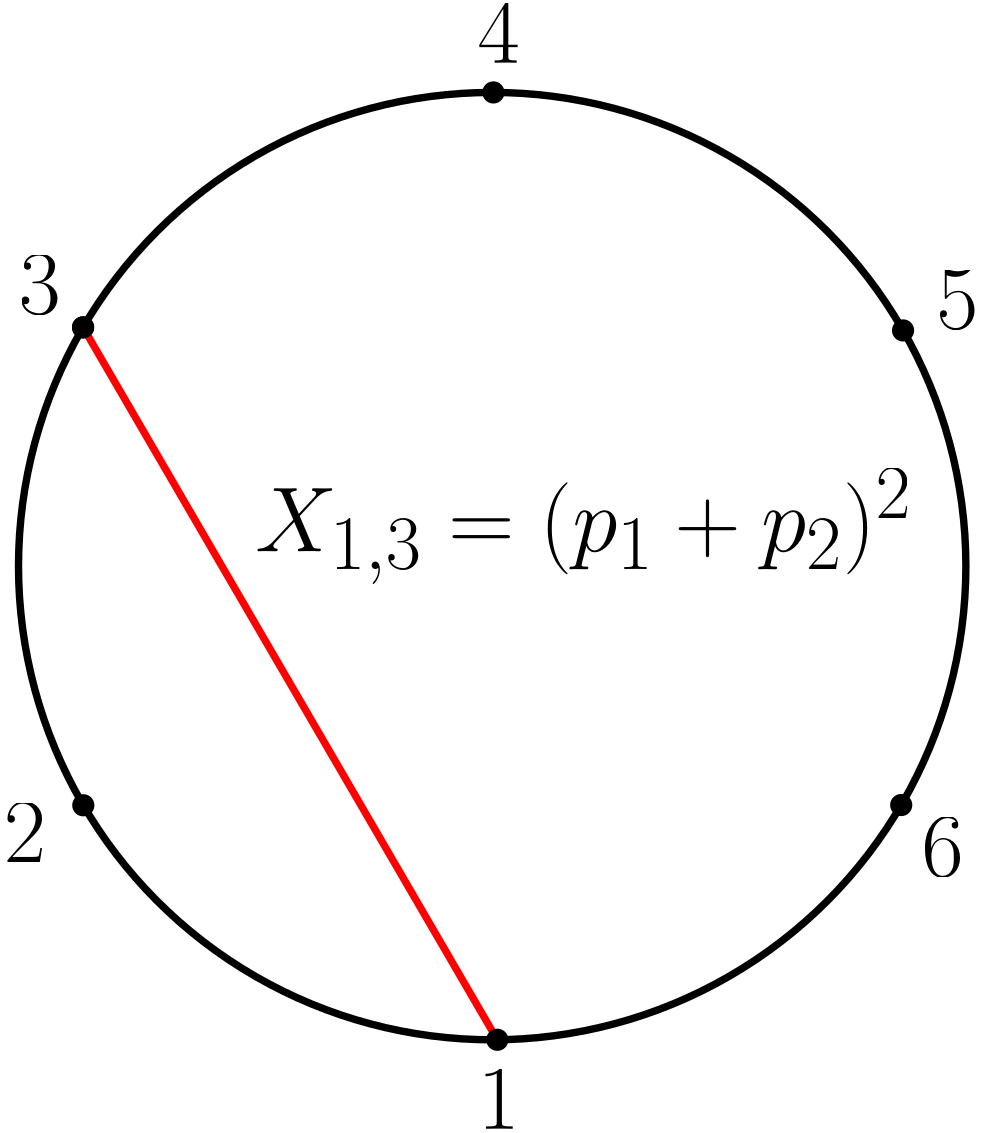}
		\caption{The 6-point surface with propagator $X_{1,3}$ in red.}
		\label{fig:n-point surface}
	\end{figure}

	Since we can decompose any Tr($\Phi^3$) amplitude into color-ordered ones, it is useful to work with planar kinematic invariants $X_{i,j}$ defined as:
	\be X_{i,j} := (p_i+p_{i+1}+\ldots+p_{j-1})^2,\quad X_{i,i+1}=0,\ee
	where subscripts are always understood mod $n$. These variables correspond to internal chords in a disk with $n$ punctures $x_i$, where the momenta of the particles are defined as $p_i=x_{i+1}-x_{i}$ and thus $X_{i,j} = (x_j - x_i)^2$ (see figure \ref{fig:n-point surface}).
	They also form a basis for all Lorentz-invariant products of the momenta. This is easily checked by noting that the number of independent $X$ variables, $n(n{-}1)/2\,{-}\,n$ coincides with the dimension of kinematic space. The $n(n{-}3)/2$ planar variables correspond precisely to the simple poles of the $n$-point $\Tr(\Phi^3)$ amplitude.
	
\begin{figure}[ht]
		\centering
		\includegraphics[width=0.3\textwidth,valign=c]{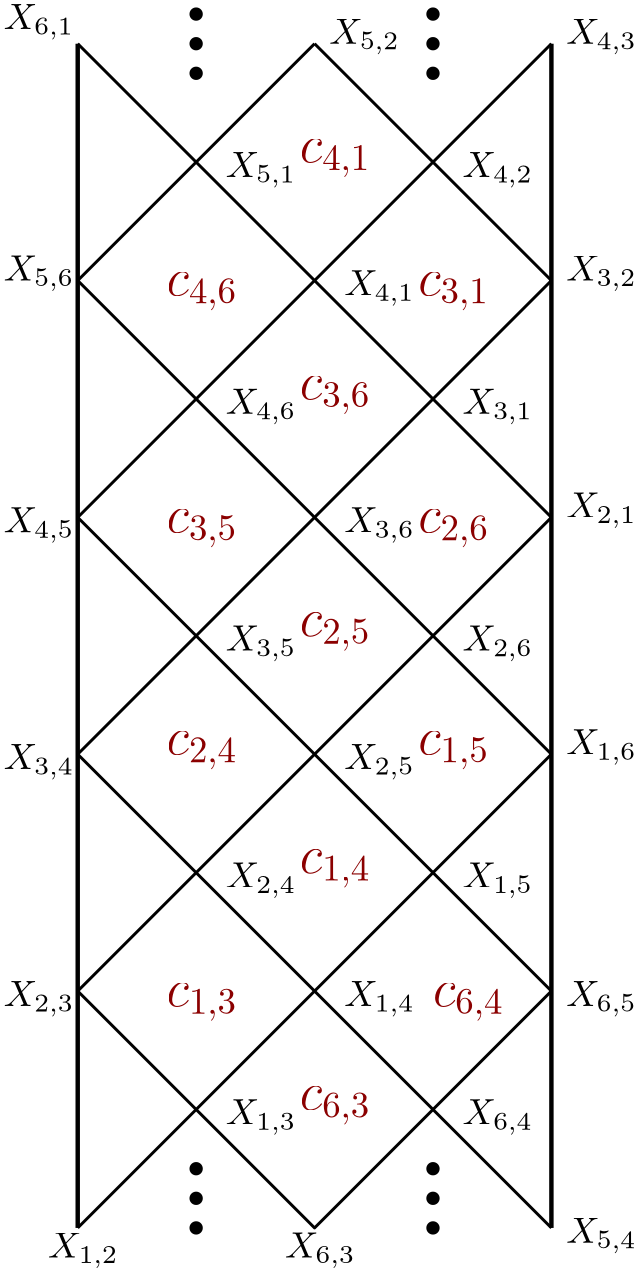}
		\caption{The 6-point kinematic mesh with $X$ variables attached to the vertices and $c$ variables to the squares.}
		\label{fig:6-pt kinematic mesh}
	\end{figure}

	Nevertheless, it is also useful to define the non-planar invariants:
	\be c_{i,j} := -2p_i\cdot p_j.\ee
	The relation between planar and non-planar variables is given by:
	\be\label{eq: c in terms of X} c_{i,j} = X_{i,j} - X_{i+1,j} - X_{i,j+1} + X_{i,j}.\ee
	This equation is elegantly encoded in the so-called \emph{kinematic mesh} \cite{Arkani-Hamed:2019vag}. It is realized as a two-dimensional infinite grid divided into diamonds (see figure \ref{fig:6-pt kinematic mesh} for an example at six points). The vertices of the diamonds are labeled by planar variables $X_{i,j}$, such that going up the left (right) edge increases the index $i$ ($j$) by one. The vanishing variables $X_{i,i+1}=0$ are placed on the vertical boundaries of the grid. The cyclic symmetry of the variables $X_{i+n,j}=X_{i,j}$ is then apparent once we make the identification $X_{i,j}=X_{j,i}$.
	It is natural to assign the non-planar invariant $c_{i,j}$ to the diamond with $X_{i,j}$ at its base, since the four $X$ variables in \eqref{eq: c in terms of X} appear on its four corners. Moreover, for any rectangular region in the mesh (called a causal diamond) the following relation holds:
	\be\label{eq:causal diamond relation} \sum_{i,j\, \in\, \diamond} c_{i,j} = X_B + X_T - X_L - X_R,\ee
	where $X_T,X_B,X_L$ and $X_R$ are the variables attached to the top, bottom, left and right vertices bounding the diamond.
	
	As mentioned before, the collection of all planar variables $X_{i,j}$ constitutes a basis of kinematic space. However, it is often useful to consider a basis that includes only a maximal set of \emph{compatible} planar variables, meaning their non-crossing chords form a full triangulation of the kinematic surface. This can be done by as follows: pick a triangulation $\mathcal{T}$ of the kinematic surface and take the $n{-}3$ compatible planar variables that appear in $\mathcal{T}$; then take all of the non-planar variables \emph{except} those corresponding to squares $c_{i-1,j-1}$ in the kinematic mesh for which $(i,j)\in\mathcal{T}$. This set of $n{-}3$ planar and $(n{-}2)(n{-}3)/2$ non-planar variables then forms a valid basis of $\mathcal{K}_n$. A collection of this type is called a \emph{kinematic basis} of $\mathcal{K}_n$.

If we choose a kinematic basis $\mathcal{B}$ containing $n{-}3$ planar variables  $\vec{X} = (X_{i_1,j_1}^\mathcal{T},\ldots,X_{i_{n-3},j_{n-3}}^\mathcal{T})$, then any planar variable $X_{i,j}$ can be expressed as a linear combination
	\begin{align}
		\label{eq:gvec}
		X_{i,j} = \sum_{c_{kl}\in\mathcal{B}} a_{i,j;k,l} c_{k,l} + \vec{g}_{i,j}\cdot \vec{X}^{\mathcal{T}}
	\end{align}
    for some collection coefficients $a_{i,j;k,l}$ and $\vec{g}_{i,j}$; the latter is called the $g$-vector of $X_{i,j}$. Note that the $g$-vectors of $X_{i,j} \in \mathcal{T}$ are simply the $n{-}3$ unit vectors $(0,\ldots,1,\ldots,0)$. We emphasize that the $g$-vectors depend on the choice of kinematic basis $\mathcal{B}$.
	
	The set of $g$-vectors of all planar variables $X_{i,j}$ form the \emph{Feynman fan}. These $g$-vectors also arise as normal vectors to the facets of the ABHY associahedron \cite{Arkani-Hamed:2017mur}, as we will review in Section \ref{sec:associahedron interpretation}, and they will be crucial in defining the deformations that are the subject of this paper.
	
	\section{Amplitudes at infinity}
	\label{sec:AatInfinity}
	In this section we define the kinematic deformations that we study in this work and summarize our main result \eqref{eq:mainresult}.
    Our motivation is to find interesting generalizations of the kinematic shift
    \be\label{eq:delta=2 shift} X_{i,j}\to\begin{cases}
    X_{i,j} + \delta,& i,j\quad \text{are both even},\\
    X_{i,j} - \delta,& i,j\quad \text{are both odd},\\
    X_{i,j}& \text{otherwise},
\end{cases}\ee
called the $\delta$-shift, that was used in \cite{Arkani-Hamed:2023swr,Arkani-Hamed:2023jry} to connect the amplitudes of $\Tr(\Phi^3)$ theory to those of the NLSM model and YM theory.
    We look for specific shifts in the $X_{i,j}$ variables that preserve enough non-planar invariants to keep a given kinematic basis intact. We can accomplish this by fixing some kinematic basis, as reviewed in Section \ref{sec: review kin space}, and then shifting each planar variable $X_{i,j}$ by an amount proportional to its $g$-vector in that basis:
	\be\label{eq: g-vector shift} X_{i,j}\to X_{i,j} + z \, \vec{g}_{i,j}\cdot\vec{t},\ee
	where $z$ is a scaling parameter with mass dimension 2 and $\vec{t}\in\mathbb{R}^{n-3} \setminus \{0\}$ specifies an arbitrary direction in the Feynman fan.  We will call \eqref{eq: g-vector shift} a \emph{$g$-vector shift}.
	
	The crucial property of $g$-vector shifts is that they automatically leave all non-planar variables in the kinematic basis unchanged, for any choice of $\vec{t}$. This is straightforward to show: start by rewriting \eqref{eq:gvec} as
 	\be \label{eq:c's invariant after shift} \sum_{c_{k,l}\in\mathcal{B}} a_{i,j;k,l}\,c_{k,l} = X_{i,j} - \vec{g}_{i,j}\cdot\vec{X}^{\mathcal{T}}.\ee
    Now under the $g$-vector shift \eqref{eq: g-vector shift}, the right-hand side becomes
    \be X_{i,j} - \vec{g}_{i,j}\cdot\vec{X}^{\mathcal{T}} \Rightarrow (X_{i,j} +z \, \vec{g}_{i,j}\cdot\vec{t}) - \vec{g}_{i,j} \cdot (\vec{X}^{\mathcal{T}} + z \vec{t})\ee
since the $g$-vector of each basis variable $X_{i_a,j_a}^{\mathcal{T}}$ is just the $a$-th unit vector. Therefore
the two contributions cancel, and we conclude that the left-hand side of \eqref{eq:c's invariant after shift} is invariant under the shift.
    	
	Performing a $g$-vector shift on a  $\Tr(\Phi^3)$ amplitude introduces dependence on $z$ and $\vec{t}$ that is complicated in general, but a relatively simple and beautiful story comes into focus if we zoom out to the ``world at infinity'' by taking the $z\to \infty$ limit for fixed $\vec{t}$. In particular, we use $\ah_n^\infty$ to denote the leading term in the large $z$ limit of the $n$-point $\Tr(\Phi^3)$ amplitude:
	\be \label{eq:leadingpower} \ah_n^{\Phi^3} \xrightarrow[z\to\infty]{\vec{t}} \frac{1}{z^k} \ah_n^\infty + \oh(z^{-k-1}),\ee
	where $k$ is determined by the details of the shift in a manner that we will discuss below, and the main analytic result of our paper is the universal formula for the leading contribution given by
	\begin{align}
		\label{eq:mainresult}
		\boxed{\ah_n^\infty \propto \sum_{\mathcal{T}_{\rm partial}} \left(\prod_{i,j\in\mathcal{T}_{\rm partial}} \frac{1}{X_{i,j}}\right)\prod_{\substack{S_i\subset S\\n_i\geq 4}} c_{S_i}}.
	\end{align}
	The remainder of this section is devoted to carefully explaining all of the ingredients in this result.  For now we note
    that all of the dependence on the planar variables $X_{i,j}$ is explicit, the $c_{S_i}$ factors depend only on the non-planar variables, and the omitted proportionality constant depends only on the choice of $\vec{t}$, as does the
    set of partial triangulations  $\mathcal{T}_{\rm partial}$ appearing in the sum.
    The formula \eqref{eq:mainresult} also admits a geometric interpretation in terms of projecting the associahedron that we will discuss in Section \ref{sec:associahedron interpretation}. Interestingly, for certain special directions $\vec{t}$ the resulting $\mathcal{A}_n^\infty$ is an actual physical scattering amplitude of a scalar-pion theory. Paired with the geometric description of the amplitude at infinity, we can construct the polytope associated to amplitudes involving two pions and an arbitrary number of scalar external states. This polytope description, one of the first of a theory with higher-derivative interactions, is presented in Section \ref{sec:polytope}.
	
	\subsection{Construction of the result}
	\label{sec:kinematic dependence}
	We now present the set of steps to calculate $g$-vector shifted $\Tr(\Phi^3)$ amplitudes to leading order in $1/z$. We focus on the six-point amplitude $\ah_6^{\Phi^3}$ as an example that illustrates the procedure.
	
\begin{figure}[ht]
		\centering
		\includegraphics[width=0.23\textwidth, valign=c]{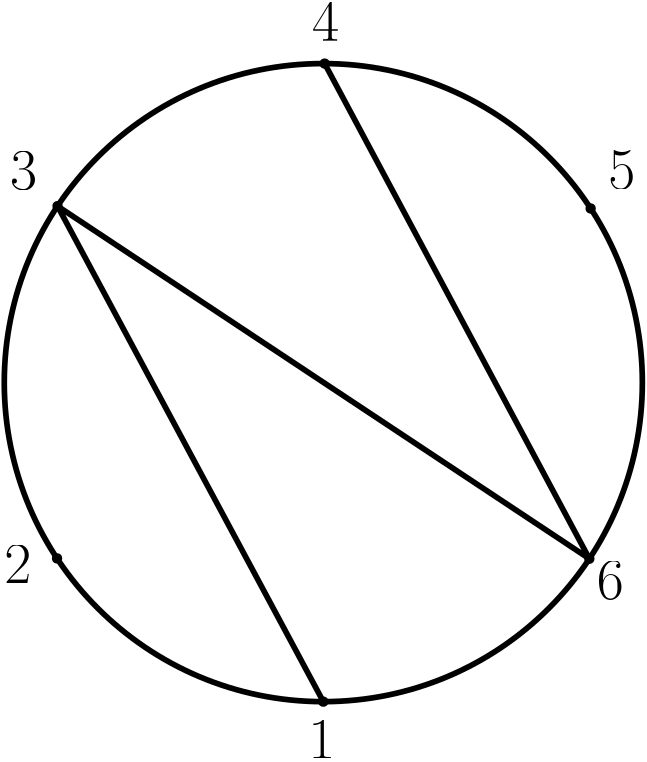}
		\caption{Zig-zag triangulation for the 6-point surface.}
		\label{fig:6-pt zig-zag triang}
	\end{figure}

	We start by choosing a kinematic basis.
	For example, we can choose a zig-zag triangulation at six points (see figure \ref{fig:6-pt zig-zag triang}) given by $\{X_{1,3},X_{3,6},X_{4,6}\}$.
	For the non-planar invariants, we discard $c_{2,6},\,c_{2,5}$ and $c_{3,5}$, leaving us with the following kinematic basis:
	\be \{X_{1,3},X_{3,6},X_{4,6},c_{1,3},c_{1,4},c_{1,5},c_{2,4},c_{3,6},c_{4,6}\}.\ee
	We can now express the rest of the planar Mandelstam variables in terms of the elements in our basis using the relation \eqref{eq:causal diamond relation}. For our case, we end up with:
	\be\label{eq: 6-pt kinematic basis} \ba X_{1,4}&=c_{3,6}+X_{1,3}-X_{3,6}+X_{4,6},\\ X_{1,5}&=c_{3,6}+c_{4,6}+X_{1,3}-X_{3,6},\\ X_{2,4}&=c_{1,3}+c_{3,6}-X_{3,6}+X_{4,6},\\ X_{2,5}&=c_{1,3}+c_{1,4}+c_{3,6}+c_{4,6}-X_{3,6},\\ X_{2,6}&=c_{1,3}+c_{1,4}+c_{1,5}-X_{1,3},\\ X_{3,5}&=c_{1,4}+c_{2,4}+c_{4,6}-X_{4,6}.\ea\ee
	The $g$-vectors read off from \eqref{eq: 6-pt kinematic basis} using \eqref{eq:gvec} are
	\begin{alignat}{3}\vec{g}_{1,3}&=(1,0,0),\quad &\vec{g}_{3,6}&=(0,1,0),\quad &\vec{g}_{4,6}&=(0,0,1),\\ \vec{g}_{1,4}&=(1,-1,1),\quad &\vec{g}_{1,5} &= (1,-1,0),\quad &\vec{g}_{2,4} &= (0,-1,1),\\ \vec{g}_{2,5} &= (0,-1,0),\quad &\vec{g}_{2,6}&=(-1,0,0),\quad &\vec{g}_{3,5}&=(0,0,-1). \end{alignat}
	A specific $g$-vector shift is then defined by a direction in the Feynman fan, given by a vector $\vec{t} \in\mathbb{R}^{n-3}$. Let us choose $\vec{t} = (t_1,t_1,t_3)$ (note that the first two components are equal). Then, using \eqref{eq: g-vector shift}, we can determine which poles are affected by the deformation and which are not. For our choice, only $X_{1,5}$ is left unshifted. Therefore, we know that the leading contribution $\ah_6^\infty$ only has poles in this variable, while the rest can appear only in the numerator.
	
	\paragraph{$\mathbf{z}$~dependence:} In general, the set of unshifted poles forms a partial triangulation $\mathcal{T}_{partial}$ that divides the kinematic surface $S$ into several subsurfaces $S_i$. We claim that the weight in $1/z$ of a specific partial triangulation is given by a product over the subsurfaces $S_i$ that contain four or more points:
	\be\label{eq: z-scaling partial triang} \mathcal{T}_{\rm partial} \sim\oh\left( \prod_{\substack{S_i\subset S\\|S_i|\geq 4}} \frac{1}{z^{|S_i|-2}}\right).\ee
	In our 6-point example, there is only one partial triangulation $\{X_{1,5}\}$, dividing the 6-point surface into two subsurfaces (1,2,3,4,5) and (1,5,6), of which only the former contributes. Altogether, the 6-point amplitude falls off as $\mathcal{O}(1/z^3)$ as $z \to \infty$.
	
	The formula \eqref{eq: z-scaling partial triang} is easy to justify: firstly, any three-point subsurface won't contribute to the $z$-scaling of the partial triangulation because it doesn't contain any chords that have been shifted (it doesn't contain any chords at all). For an $n_i$-point subsurface $S_i$ (with $n_i\geq4)$, all the chords one can draw inside it have been shifted. The coefficient in front of the unshifted propagators $1/X_{i,j}^{\mathcal{T}_{\rm  partial}}$ for the subsurface $S_i$ is given by the shifted lower-point amplitude of this subsurface. Naively, since we shift all the chords, we would expect the leading term to scale as $\oh(1/z^{n_i{-}3})$. However, amplitudes in Tr($\Phi^3$) are known to scale as $1/z^{n{-}2}$ \cite{Arkani-Hamed:2017mur}; this is linked to the projective invariance of the scattering form which we comment on more in Section \ref{sec:introassoc}. This scaling of the lower-point amplitude then implies that the leading term scales as $\oh(1/z^{|S_i|{-}2})$. Repeating this reasoning for all subsurfaces created by $\mathcal{T}_{\rm partial}$, we reach the result \eqref{eq: z-scaling partial triang}. This is a generalization of the argument presented in \cite{He:2018svj,Yang:2019esm}, where the faster-than-expected fall-off of the shifted amplitude was used to derive recursion relations.
	
\begin{figure}[ht]
		\centering
		\includegraphics[width=0.6\textwidth,valign=c]{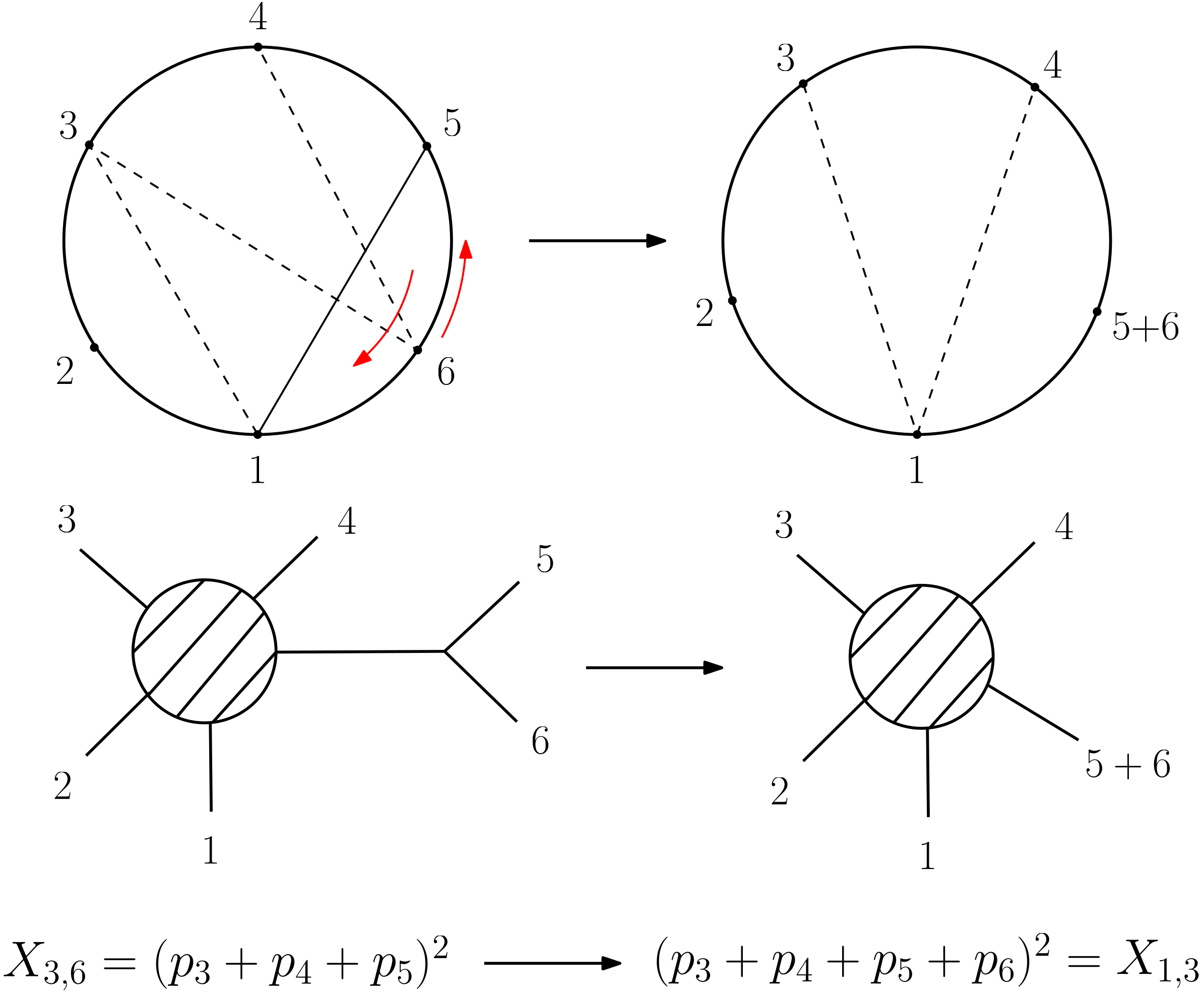}
		\caption{Top: Kinematics and base triangulation inherited by the off-shell subsurface (1,2,3,4,5). Bottom: Interpretation in terms of Feynman diagrams with massive legs.}
		\label{fig:5-pt subsurface}
	\end{figure}

	For example at 6-point, there is only one type of Feynman diagram that contributes (see figure \ref{fig:5-pt subsurface}). Naively one might have expected these to scale as the number of missing propagators, i.e. as $1/z^2$, yet we find that this scaling is enhanced to $1/z^3$, in agreement with our general argument above.

   In some cases, we can predict the $z$ behavior without a careful consideration of subsurfaces: these are the directions for which one or more of the entries of
    \be \vec{v} = ( t_1, t_2 - t_1, t_3 - t_2, \cdots, t_{n-3} - t_{n-4}, t_{n-3})\ee
    are zero.  We note that upon taking the limit $1/z \to \infty$ along this direction, the leading term in the shifted amplitude scales simply as $z^{n{-}2 {-} q}$ where $q$ is the number of entries of $\vec{v}$ that are zero.
	
	\paragraph{Kinematic dependence:} To determine the shifted amplitude, we start with the set of all partial triangulations that have the lowest weight in $1/z$, since we're interested in the leading term $\ah_n^\infty$. Next, focus on one particular partial triangulation made of unshifted poles, and consider each subsurface $S_i$ with four or more points separately. Associated to any $S_i$ and a triangulation thereof, there exists a contact term $c_{S_i}$ which is a function of the non-planar $c_{i,j}$'s. We now explain how to extract the contact term of a generic surface, and then use this result to determine the numerator for a specific partial triangulation.
	
	Given a canonically ordered surface $(1,2,\ldots, k)$ triangulated by a subset of $X_{i,j}^\mathcal{T}$ and its corresponding kinematic mesh, draw all the maximal causal diamonds where each $X_{i,j}^\mathcal{T}$ is located at the bottom vertex. The contact term is given by the square(s) where all causal diamonds overlap simultaneously. Lastly, there is a sign associated to each $c_{i,j}$, given by $(-1)^{i+j+1}$. Note that this means the numerator can be written purely in terms of the relevant $c_{i,j}$'s.
		
	In order to use this method to determine the contact term for each of the subsurfaces $S_i$, we need them to inherit the kinematics and base triangulation from the bigger surface $S$. This is done by rotating all the points outside $S_i$ counterclockwise and all the chords clockwise until they are included inside $S_i$. In our example, the subsurface (1,2,3,4,5) will inherit the information about the whole (1,2,3,4,5,6) surface by rotating the point 6 counterclockwise and the chords $X_{3,6},X_{4,6}$ clockwise (see figure \ref{fig:5-pt subsurface}, top).

	This procedure is justified very naturally from a Feynman diagram point of view. The subsurface $S_i$ can also be thought of as a diagram where some of the external legs corresponding to internal chords $X_{i,j}^{\mathcal{T}_{\rm partial}}$ are massive, with momenta equal to $p_i+\ldots+p_{j-1}$. Meanwhile, any propagator from the base triangulation $X_{i,j}^{\mathcal{T}_{\rm base}}$ that is not compatible with the one from the partial triangulation will be \emph{moved into} $S_i$ by taking into account the legs that have become massive. This effectively amounts to rotating points counterclockwise and chords clockwise in the surface (see figure \ref{fig:5-pt subsurface}, bottom). Once the subsurface $S_i$ has a base triangulation, we can assign a contact term to it using the procedure outlined above.
	
\begin{figure}[ht]
		\centering
		\includegraphics[width=0.3\textwidth,valign=c]{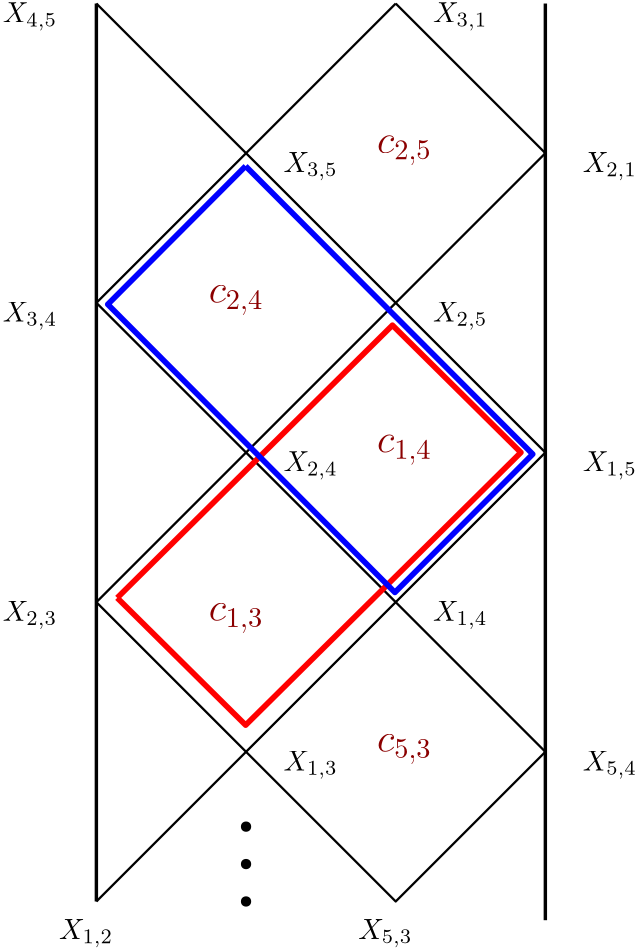}
		\caption{Maximal causal diamonds with $X_{1,3}$ (red) and $X_{1,4}$ (blue) at their base. They overlap on the square corresponding to $c_{1,4}$.}
		\label{fig:5-pt max causal diamonds}
	\end{figure}

	Let us return to our example. For the subsurface (1,2,3,4,5), our inherited base triangulation is $\{X_{1,3},X_{1,4}\}$. If we draw the 5-point kinematic mesh with two maximal causal diamonds with those planar variables at the bottom, they both overlap in the square associated to $c_{1,4}$ (see figure \ref{fig:5-pt max causal diamonds}). Thus, the contact term of this subsurface is given by the non-planar variable $c_{1,4}$ (with positive sign). Meanwhile, the other subsurface (5,6,1) has only three points and so has a trivial contact term of 1.
	
	Finally, the total contribution to $\ah_n^\infty$ is given by a sum over all these lowest weight partial triangulations $\mathcal{T}_{\rm partial}$, including the poles and the numerators as described above, leading thereby to the main result advertised in \eqref{eq:mainresult}.

	For our running example, we have a single partial triangulation with the chord $X_{1,5}$, for which we have already determined the numerator to be  $c_{1,4}$. The shifted amplitude is then just
	\be \ah_6^\infty \propto \frac{c_{1,4}}{X_{1,5}}.\ee
	
	\paragraph{$\vec{t}$ dependence:} When performing a generic shift $\vec{t}\in\mathbb{R}^{n-3} \setminus \{0\}$, the leading contribution to the amplitude $\ah_n^\infty$ is composed of a factor depending only on the kinematics (the computation of which we have described in the previous subsection) and a factor depending only on the free components of $\vec{t}$. The latter is arranged in a Parke-Taylor-like form that also turns out to factorize in terms of the subsurfaces that result from the partial triangulations made of unshifted poles.
	
	More specifically, when performing a totally generic shift $\vec{t}=(t_1,t_2,\ldots,t_{n-3})$ on the $n$-point scalar amplitude $\ah_n^{\Phi^3}$, we observe the following result:
	\be \ah_n^{\Phi^3} \xrightarrow[z\to\infty]{\vec{t}=(t_1,t_2,\ldots,t_{n-3})} \frac{1}{z^{n-2}} \frac{c_S}{t_1(t_1-t_2)(t_2-t_3)\cdots (t_{n-4}-t_{n-3})t_{n-3}} = \frac{1}{z^{n-2}}\frac{c_S}{\text{PT}_S},\label{eq:tdependence} \ee
	 where $c_S$ is the contact term of the whole surface $(1,2,\ldots,n)$ for the chosen triangulation. We can then identify the product of $t_i$'s in the denominator as the Parke-Taylor factor for the $n$-point surface, $\text{PT}_S$. Here we see explicitly that directions in which $t_{i{+}1}=t_i$ are special.
    
	This behavior is also exhibited by general shifted amplitudes. In order to precisely determine the $t$-dependence for a shift, we simply look at the partial triangulations with the minimal number of chords that contribute to leading order in the large $z$ expansion. The denominator of $\ah_n^\infty$ then includes a product of the Parke-Taylor factors of the corresponding subsurfaces, where the base triangulation is inherited in the way explained in Section \ref{sec:kinematic dependence}:
	\be \ah_n^\infty \propto \prod_{S_i \subset S} \frac{1}{\text{PT}_{S_i}}.\ee
	The fact that all the partial triangulations are assigned the same Parke-Taylor factor is not really surprising. Indeed, this is just a consequence of the fact that all $\mathcal{T}_{\rm partial}$ only include unshifted chords, whose $g$-vectors are perpendicular to the shift direction $\vec{t}$. Since the $t_i$ terms of the shifted amplitude only depend on this direction, all partial triangulation will show the same behavior.
	
	To illustrate this with an example, consider the same six-point shift $\vec{t}=(t_1,t_1,t_3)$ in the zig-zag triangulation that we worked with in the last section. As we saw previously, the leading contribution to $\ah_6^\infty$ is given by a single partial triangulation with the chord $X_{1,5}$. From the resulting five-point subsurface, we obtain the following Parke-Taylor factor:
	\be \text{PT}_S = t_1(t_1-t_3)t_3,\ \text{where}\ S=(1,2,3,4,5+6).\ee
	The other three-point subsurface doesn't have any $t$-dependence. With this, we are finally able to write the complete expression for the leading order of the shifted amplitude:
	\be \ah_6^{\Phi^3}\xrightarrow[z\to\infty]{\vec{t}=(t_1,t_1,t_3)} \frac{1}{z^3}\times\frac{1}{t_1(t_1-t_3)t_3}\times\frac{c_{1,4}}{X_{1,5}}\,,\ee
	which is consistent with \eqref{eq:mainresult}.

	\subsection{Factorization and commutativity at infinity}\label{sec:factorization}
	
	We now remark on an interesting property of our $g$-vector shifts: the shifted amplitudes present a completely consistent factorization behavior onto themselves.  In other words, the world at infinity is closed under factorization. Indeed, since $\ah_n^\infty$ is a sum over partial triangulations with unshifted chords, the residue on any one of them will result in a direct product of lower-point objects:
	\be \Res_{X=0} \ah^\infty_n = \ah^\text{cut}_{m}\times \ah^\text{cut}_{n-m+2},\ee
	where the cut amplitudes $\ah_m^\text{cut}$ only contain the poles associated to unshifted chords of the corresponding subsurface $S_m^\text{cut}$, weighed by the appropriate numerator factors $\prod c_{S_i}$. However, that is precisely a lower-point shifted amplitude itself. In other words:
	\be \ah^\text{cut}_{m} = \ah_m^\infty,\ee
	making the shifts manifestly consistent with factorization. Similar factorization properties were discussed in \cite{Jagadale:2021iab} in the strict $z=\infty$ limit (this would be 0 in the cases we discuss)\footnote{In these cases, a large deformation results in a positive geometry called an accordiahedron \cite{Jagadale:2021iab} that inherits factorization behavior from the associahedron.}. Indeed while $g$-vector shifts are not the only ones that preserve factorization properties, they are the only families of shifts where the combinatorial structure of the lower-point subsurfaces is preserved. This is what allows us to completely determine the numerator structure for the shifted objects.
	
	Another interesting property is commutativity: applying two different shifts given by $\vec{t}_1, \vec{t}_2$ to the scalar amplitude $\ah_n^{\Phi^3}$ yields the same result regardless of the order:
	\be \underset{z\to\infty}{\lim_{X\to X+\vec{g}\cdot \vec{t}_2 z}}\ \ \underset{z\to\infty}{\lim_{X\to X+\vec{g}\cdot \vec{t}_1 z}} \ah^{\Phi^3}_n = \underset{z\to\infty}{\lim_{X\to X+\vec{g}\cdot \vec{t}_1 z}} \ \ \underset{z\to\infty}{\lim_{X\to X+\vec{g}\cdot \vec{t}_2 z}}\ah^{\Phi^3}_n\,.\ee
	Here we mean on each side the leading non-vanishing term after both limits are taken. This happens because the set of preserved chords and hence the subsurfaces formed by these chords are insensitive to the order of limits, making large $g$-vector shifts a completely ``Abelian'' operation on the surface itself.
	
	\subsection{Pion behavior in shifted particles}\label{sec: pion behavior}
	
	While we have seen that the objects living at infinity do satisfy factorization, they are not in general amplitudes of any discernible theory. In this section, we discuss special cases of $g$-vector shifts in which we can identify the results as known amplitudes for tree-level processes involving pions and/or scalar particles. Consider for example a $g$-vector shift at $n=4$. There is only one possible direction $\vec{t}=(1)$, which translates to $X_{1,3}\to X_{1,3}+z$ and $X_{2,4}\to X_{2,4}-z$. This yields:
	\be \ah_4^{\Phi^3} = \frac{1}{X_{1,3}} + \frac{1}{X_{2,4}} \to -\frac{c_{1,3}}{z^2} + \oh(z^{-3}),\ee
	where we have used $X_{1,3}+X_{2,4}=c_{1,3}$. Although very simple, this result already illustrates that the residual contact term left after shifting is the four-point pion amplitude in a NLSM. This is not a surprise: the kinematic shift is proportional to the $\delta$ shift of the stringy integral described in \cite{Arkani-Hamed:2023swr}, which in the low-energy limit $\alpha'\to0$ (equivalently, the large-$z$ limit) results in the four-point pion amplitude.
	
	At $n=5$, we know that we will not be able to identify $\mathcal{A}_n^\infty$ as a pion amplitude, simply because of the fact that we started with an odd-point process and pions in NLSM only have even-point interactions. Nevertheless, let us consider for example $\vec{t}=(0,1)$:
	\be\label{eq:5-pt (0,1) shift X's}\ba X_{1,3}&\to X_{1,3},\\ X_{1,4}&\to X_{1,4}+ z,\\ X_{2,4}&\to X_{2,4}+z,\\ X_{2,5}&\to X_{2,5},\\ X_{3,5}&\to X_{3,5} - z.\ea\ee
	Using \eqref{eq:mainresult}, we get
	\be\label{eq:5-pt (0,1) shift amp} \ah_5^{\Phi^3}\xrightarrow[z\to\infty]{\vec{t}=(0,1)} -\frac{1}{z^2}\left(\frac{c_{1,4}+c_{2,4}}{X_{1,3}} + \frac{c_{1,4}}{X_{2,5}}\right).\ee
	The result above can be identified as the mixed amplitude $\ah_5[\phi_1,\phi_2,\pi_3,\pi_4,\phi_5]$, where three scalar particles interact with two pions.
	
	However, this is not always the case. Consider for instance the example we introduced in Section \ref{sec:kinematic dependence}. Taking the 6-point amplitude expressed in the basis associated to the zig-zag triangulation and performing a $g$-vector shift in the direction given by $\vec{t}=(t_1,t_1,t_3)$, we obtain:
	\be \ah_6^{\Phi^3}\xrightarrow[z\to\infty]{\vec{t}=(t_1,t_1,t_3)} \frac{1}{z^3}\times\frac{1}{t_1(t_1-t_3)t_3}\times\frac{c_{1,4}}{X_{1,5}}.\ee
	It is clear from the kinematic dependence that this cannot be recognized as a mixed amplitude between scalars and pions. Indeed, from the appearance of the collinear $X_{1,5}$ pole, we deduce that particles 5 and 6 must still behave as scalars with a cubic coupling. This leaves us with an effective five-point process (1,2,3,4,I), where $I=(5{+}6)$. At leading order in the cubic scalar coupling, the only possibility would be that three of the particles were scalars, while the other two behaved as pions. However, this interpretation would require another two-particle pole (e.g. $X_{1,3}$ or $X_{2,4}$), which we don't see in the shifted object. We thus conclude that this shift does not give rise to a familiar mixed amplitude\footnote{One can always build a Lagrangian which generates scattering amplitudes matching any kinematic dependence as long as the number of scalar species is large enough. In this example, the amplitude could be $\mathcal{A}_6[\phi_1,\phi_2,\phi_3,\phi_4,\phi_5,\phi_5]$ and the corresponding Lagrangian could have two terms $\phi_5^3$ and $\partial^\mu\phi_1\phi_2\phi_3\partial_\mu\phi_4\phi_5$. Here the subscripts are flavor indices.}.
	
	Despite not being able to consistently identify the shifted objects as well-defined amplitudes, we can still characterize what type of behavior each individual particle shows after the shift. In particular, we can unequivocally state that given an expression describing some scattering process, a given particle behaves as a pion as long as it doesn't participate in any two-particle pole and the Adler zero \cite{Adler:1964um} is satisfied. The latter condition, also known as the Weinberg soft pion theorem, is stated as:
	\begin{align}
		\label{eq:Adler}
		\mathcal{A}_n\sim \mathcal{O}(\tau)
        \end{align}
        when the momentum $p_i = \tau \hat{p}_i$ of a pion is taken to zero as $\tau \to 0$ with $\hat{p}_i$ fixed. 
	Amplitudes in Tr($\Phi^3$) theory were shown to vanish when $c_{i,j}=0$ for all $j$ non-adjacent to $i$ in \cite{Arkani-Hamed:2023swr}. Further discussion of these zeros and their near-zero splitting can be found in \cite{Bartsch:2024amu, Li:2024qfp, Zhang:2024iun, Zhou:2024ddy, Li:2024bwq, Zhang:2024efe}. This is known as the \emph{skinny rectangle zero}, since the corresponding set of squares in the kinematic mesh form a rectangle of width one. In the absence of poles in $X_{j,j{+}2}=0$ and $X_{j,j{-}2}=0$, the skinny rectangle zero implies the presence of the Adler zero \cite{Adler:1964um}. In addition, it is known that pion interactions are fully fixed in terms of their Adler zeros \cite{Cheung:2014dqa, Cheung:2015ota}.
		
	These two conditions -- the absence of collinear poles and the presence of a skinny rectangle zero -- are completely straightforward to check for a certain $g$-vector shift, even without knowing the precise expression for the final object. For the former, we can determine which two-particle poles are going to be removed and which are being preserved by the shift. For the latter, we know that a $g$-vector shift preserves all non-planar invariants $c_{i,j}$ that are included in the basis. Thus, we only need to check whether the ones outside of the kinematic basis are preserved. This can easily be done using \eqref{eq: c in terms of X}.
	
	Let us return to our example above. After shifting in the $\vec{t}=(t_1,t_1,t_3)$ direction, the only pole that is preserved from the scalar amplitude is $X_{1,5}$. As before, we deduce that particles 5 and 6 cannot behave as pions. As for the skinny rectangle zero, we look at the non-planar variables outside of the kinematic basis given by the zig-zag triangulation, which are $c_{2,4},\, c_{2,6}$ and $c_{3,5}$. Using \eqref{eq: c in terms of X}, we see that:
	\be \ba
	c_{2,4} &= X_{2,4} - X_{2,5} + X_{3,5} \xrightarrow[z\to\infty]{\vec{t}=(t_1,t_1,t_3)} c_{2,4},\\ c_{2,6} &= X_{2,6} - X_{3,6} + X_{1,3} \xrightarrow[z\to\infty]{\vec{t}=(t_1,t_1,t_3)} c_{2,6} - t_1 z,\\
	c_{3,5} &= X_{3,5} - X_{3,6} + X_{4,6} \xrightarrow[z\to\infty]{\vec{t}=(t_1,t_1,t_3)} c_{3,5} -t_1 z.\ea\ee
	Since $c_{2,6}$ and $c_{3,5}$ are shifted, we conclude that particles 2 and 3 will not behave as pions at infinity, as the amplitude won't satisfy the requisite Adler zero condition. This leaves us with particles 1 and 4, which do satisfy both constraints and thus do behave as pions in the shifted amplitude at infinity.
	
\section{Projecting the ABHY associahedron}
\label{sec:associahedron interpretation}

Amplitudes in Tr($\Phi^3$) theory are canonical functions on polytopes called ABHY associahedra \cite{Arkani-Hamed:2017mur}. In this section, we study the effect of large $g$-vector shifts on the polytope description of $\ah_n^\infty$.
		
	\subsection{Introduction to the ABHY associahedron}	
    \label{sec:introassoc}
	The $n$-particle \emph{ABHY associahedron} $A_n$ encodes the whole combinatorial structure of the sum over all possible tree-level Feynman diagrams for a fixed number $n$ of external particles \cite{Arkani-Hamed:2017mur}. Here we review its construction. We begin with two regions in kinematic space: the first is the simplex $\Delta_n$ restricted by
	\be X_{i,j} > 0, \quad 1\leq i < j \leq n.\ee
	The second one is the $(n{-}3)$-dimensional subspace $H_n\subset\mathcal{K}_n$ defined by picking a triangulation $\mathcal{T}$ and imposing that all of the non-planar variables $c_{i,j}^\mathcal{T}$ variables in the associated  kinematic basis (as reviewed in Section \ref{sec: review kin space}) are positive constants.
	
\begin{figure}[ht]
		\centering
		\includegraphics[width=0.55\textwidth,valign=c]{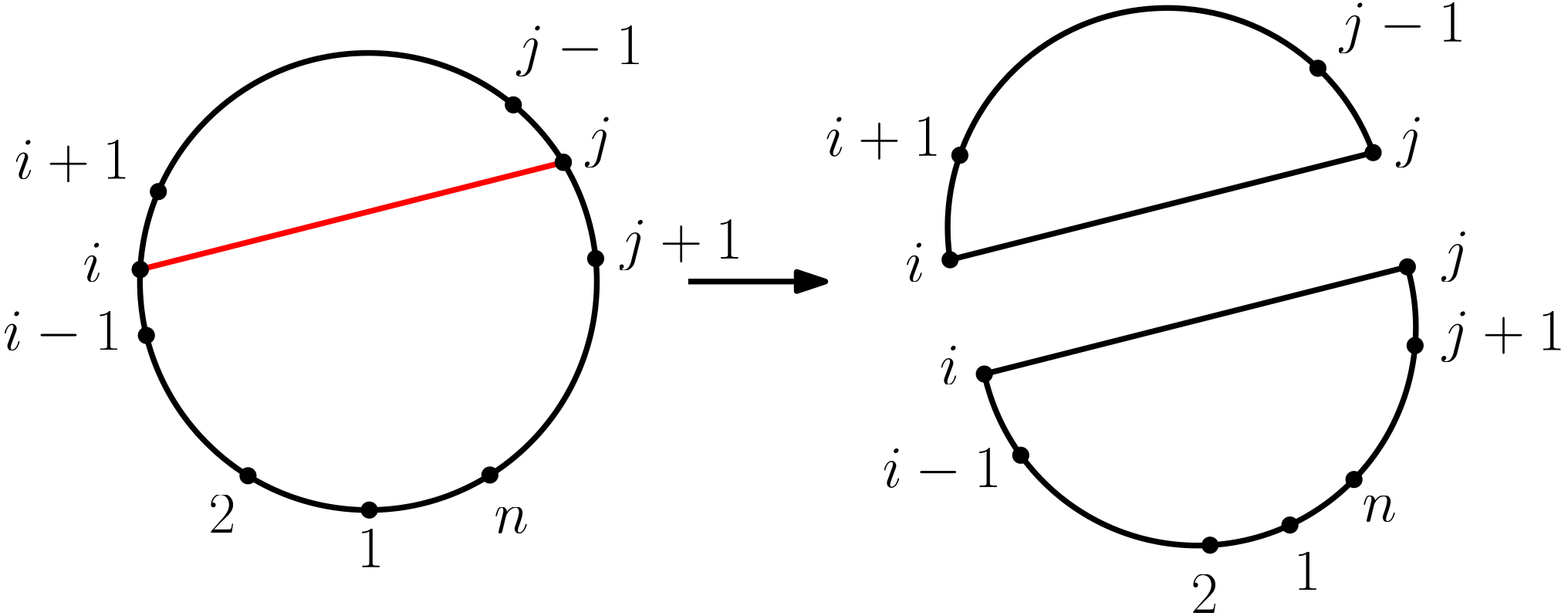}
		\caption{Factorization of the surface along the chord $X_{i,j}$}
		\label{fig:associahedron factorization}
	\end{figure}

	The kinematic associahedron is then the polytope $A_n:= H_n\cap \Delta_n$. We can embed this object in the subspace spanned by the planar variables $X_{i,j}^\mathcal{T}$ in the kinematic basis. As we'll see in the examples below, the resulting geometry is a polytope where each codimension-$d$ boundary corresponds to a partial triangulation of the kinematic surface with $d$ diagonals. Moreover, the associahedron factorizes combinatorially: any codimension-1 boundary (called a \emph{facet}) $\mathcal{F}$ associated to a chord $X_{i,j}$ in the surface is identical to the direct product of the lower-point associahedra which result from cutting the surface along $X_{i,j}$ (see figure \ref{fig:associahedron factorization}):
	\be \mathcal{F} \simeq A_m\times A_{n-m+2}.\ee
	As with any other positive geometry, there is a canonical form $\Omega_n \equiv \Omega(A_n)$ linked to the associahedron, such that the residue of $\Omega_n$ at any boundary $B$ is given by $\Omega(B)$, making the factorization properties manifest. In fact, the good factorization behavior of $\mathcal{A}_n^\infty$ discussed in Section \ref{sec:factorization} can be interpreted geometrically as the fact that when deforming the kinematics we are only moving within the $H_n$ subspace defining the associahedron in the specific basis. As a consequence, we preserve the underlying factorization structure of $\Tr(\Phi^3)$ amplitudes.

	This is related to the so-called planar scattering form, which is defined as a sum over planar cubic graphs $\mathcal{G}$, each weighed by a $d$log form:
	\be \Omega_n = \sum_{\text{planar}\ \mathcal{G}} \text{sign}(\mathcal{G})\bigwedge_{a=1}^{n-3}d\log X_{i_a,j_a},\ee
	where $X_{i_a,j_a}$ are the propagators appearing in $\mathcal{G}$ and $\text{sign}(\mathcal{G})$ depends on the ordering of the $X_{i,j}$ in the wedge product. These signs are uniquely fixed by imposing that the scattering form is projective invariant, i.e.~doesn't change under $\text{GL}(1)$ transformations $X_{i,j}\to \Lambda(X)X_{i,j}$ \cite{Arkani-Hamed:2017mur}. As we saw in Section \ref{sec:kinematic dependence}, this reflects the fact that spurious poles introduced by the Feynman diagram expansion cancel in the full amplitude.
		
	The canonical form of the associahedron itself is then defined as the pullback of the scattering form $\Omega_{n}$ to the subspace $H_n$, and determines the $n$-point tree-level amplitude for $\Tr(\Phi^3)$ theory:
	\be \Omega_n = \left(\sum_{\text{planar}\, \mathcal{G}}\ \prod_{a=1}^{n-3}\frac{1}{X_{i_a,j_a}}\right) d^{n-3}X = \ah^{\Phi^3}_n\, d^{n-3}X.\ee

	Lastly, we review the relation to the Feynman fan of $g$-vectors. Given that the associahedron is a convex polytope, it can be described as the convex hull of its facets, which are carved out by a set of equalities $Y\cdot W_i=0$, where $Y=(1,X)\in \mathbb{P}^{n-3}$ is a point in projective space and the $W_i$ are vectors in dual projective space associated to each facet. For the associahedron, these vectors $W_i$ are fixed by expressing all of the $X_{i,j}$ in terms of the elements of a kinematic basis associated with some triangulation of the surface. Note that the projection of $W_i$ onto $\mathbb{R}^{n-3}$ is precisely the definition of a $g$-vector \eqref{eq:gvec}; thus these are exactly the set of rays which form the Feynman fan and the basis of the kinematic shifts we studied in Section \ref{sec:AatInfinity}.
	
\subsection{\texorpdfstring{$n=5$ associahedron}{n=5 associahedron}}
	\begin{figure}[ht]
		\centering
		\includegraphics[width=0.25\textwidth,valign=c]{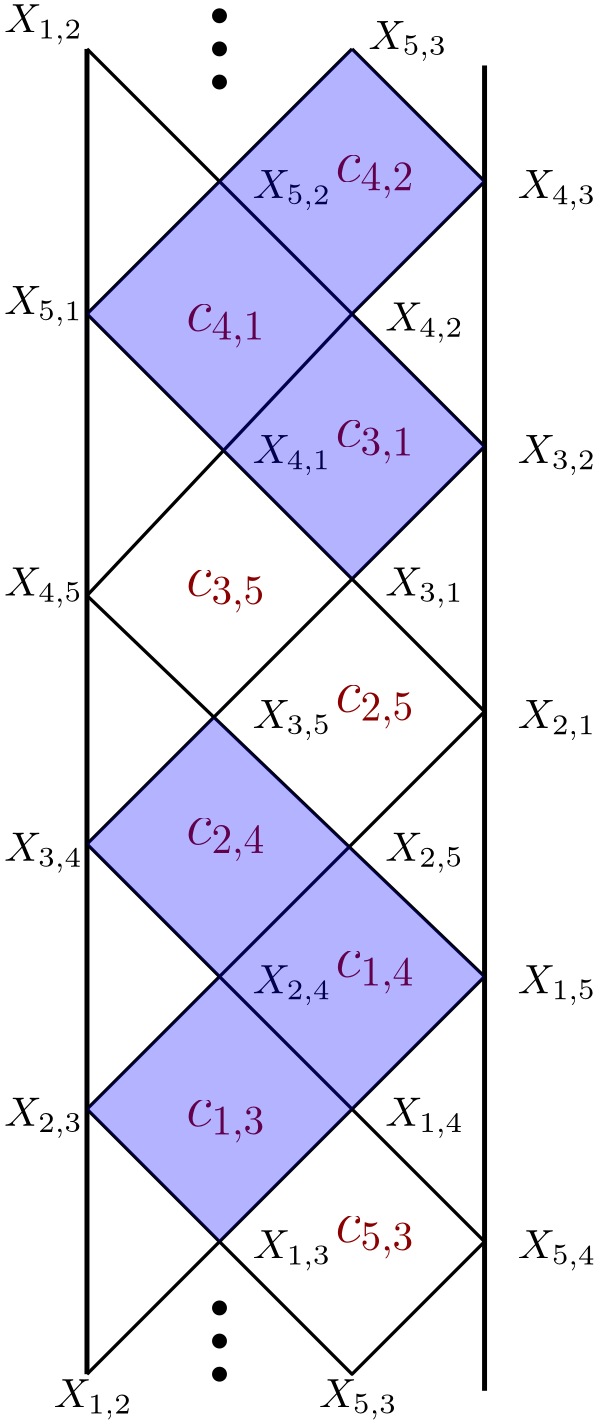}
		\caption{The five-point kinematic mesh. The shaded squares correspond to the variables of basis $\{X_{13},X_{14}\}$.}
		\label{fig: 5-pt kinematic mesh}
	\end{figure}
\begin{figure}[ht]
		\centering
		\includegraphics[width=0.8\textwidth,valign=c]{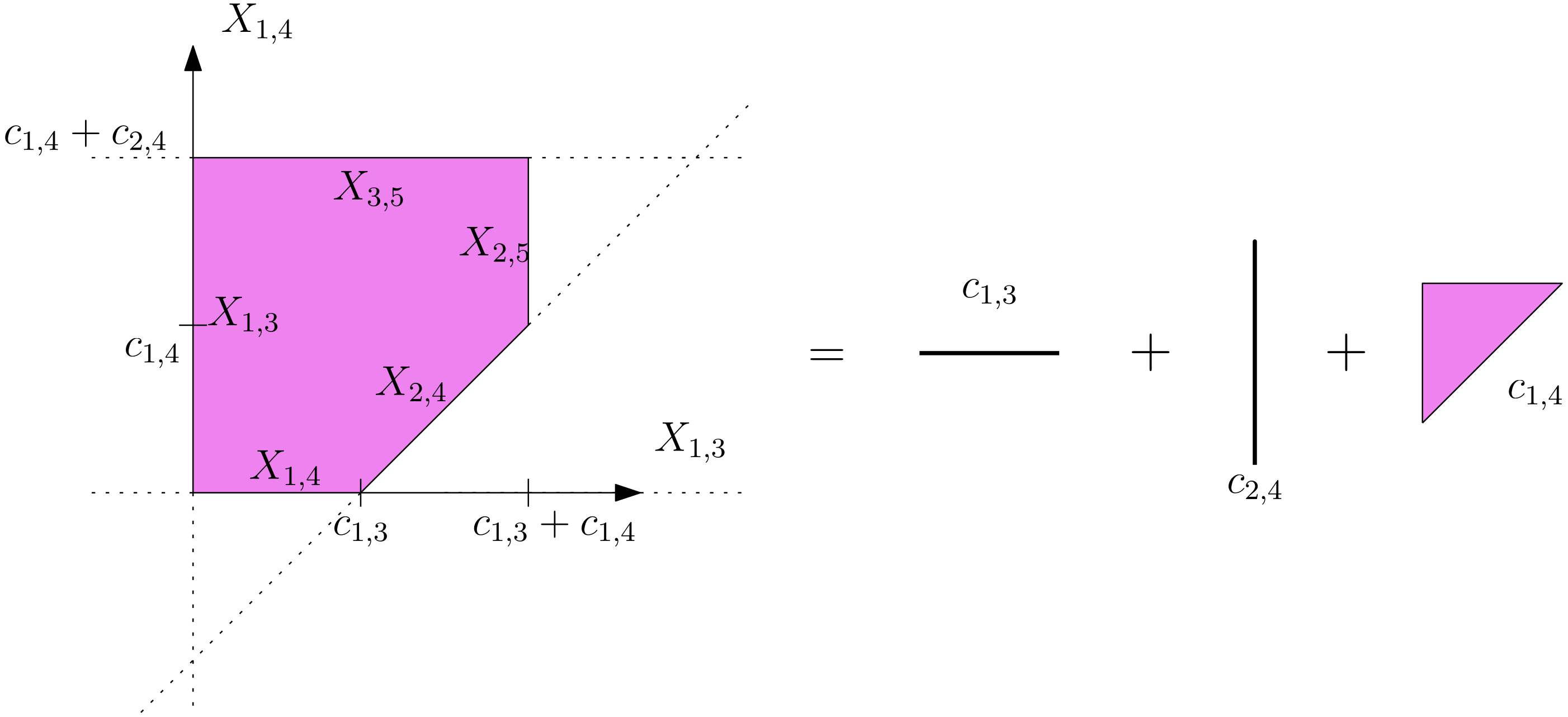}
		\caption{The five-point associahedron in the $\{X_{13},X_{14}\}$ basis and its Minkowski summands.}
		\label{fig:5-pt associahedron}
	\end{figure}

	The five-point kinematic space $\mathcal{K}_5$ is spanned by five independent variables. Choosing for instance $X_{1,3},X_{1,4}$ as triangulating chords, the non-planar variables in the basis are $c_{1,3},c_{1,4}$ and $c_{2,4}$ (see shaded region of the mesh in figure \ref{fig: 5-pt kinematic mesh}). In this basis, all other planar invariants are given by:
	\be\label{eq: 5-pt facets}\ba X_{2,4} &= c_{1,3} - X_{1,3} + X_{1,4},\\ X_{2,5}&= c_{1,3}+c_{1,4} - X_{1,3},\\ X_{3,5} &= c_{1,4} + c_{2,4} - X_{1,4}.\ea\ee
	The intersection of the simplex $\Delta_5$ defined by $X_{i,j}>0$ and the subspace $H_5$ of constant positive $c_{1,3},c_{1,4},c_{2,4}$ is given by the following inequalities in $(X_{1,3},X_{1,4})$-space:
	\be\ba &0< X_{1,3} < c_{1,3}+c_{1,4},\\ &0<X_{1,4}<c_{1,4}+c_{2,4},\\ &X_{1,3} - X_{1,4} < c_{1,3}.\ea\ee
	This carves out a pentagon as in figure \ref{fig:5-pt associahedron}.
	Importantly, we see that every codimension-1 boundary is a one-dimensional associahedron (a line segment), which encodes the combinatorics of the four-point subsurfaces that are defined by cutting along each one of the chords $X_{i,j}$. For instance, the boundary $X_{2,4}$ is a line segment with vertices associated to $X_{1,4}$ and $X_{2,5}$. These are precisely the two chords compatible with $X_{2,4}$ in the five-point surface.
	
	This behavior is also reflected in the planar scattering form:
	\be\label{eq:5-pt scattering form} \ba \Omega_5 = \frac{dX_{1,3}\wedge dX_{1,4}}{X_{1,3}X_{1,4}} + \frac{dX_{1,4}\wedge dX_{2,4}}{X_{1,4}X_{2,4}} + \frac{dX_{2,4}\wedge dX_{2,5}}{X_{2,4}X_{2,5}} + \frac{dX_{2,5}\wedge dX_{3,5}}{X_{2,5}X_{3,5}} + \frac{dX_{3,5}\wedge dX_{1,3}}{X_{3,5}X_{1,3}}\\= d\log\left(\frac{X_{1,3}}{X_{2,4}}\right)\wedge d\log\left(\frac{X_{1,4}}{X_{3,5}}\right)+d\log\left( \frac{X_{2,4}}{X_{2,5}} \right)\wedge d\log\left(\frac{X_{2,5}}{X_{3,5}}\right),\ea\ee
	where projective invariance is realized by writing $\Omega_5$ in terms of ratios of Mandelstam variables.
	
	Another important feature to note is that the associahedron can also be expressed as a Minkowski sum of simplices, each connected to a square in the kinematic mesh. For example, in this basis the pentagon in figure \ref{fig:5-pt associahedron} is a Minkowski sum of a horizontal line segment, a vertical line segment and a triangle associated to $c_{1,3}$, $c_{2,4}$ and $c_{1,4}$, respectively (this can be checked by setting the other two variables to zero and seeing what the pentagon collapses into). This will be crucial in order to determine the numerator structure of the shifted amplitudes in a geometrical way.
	
\begin{figure}[ht]
		\centering
		\includegraphics[width=0.30\textwidth,valign=c]{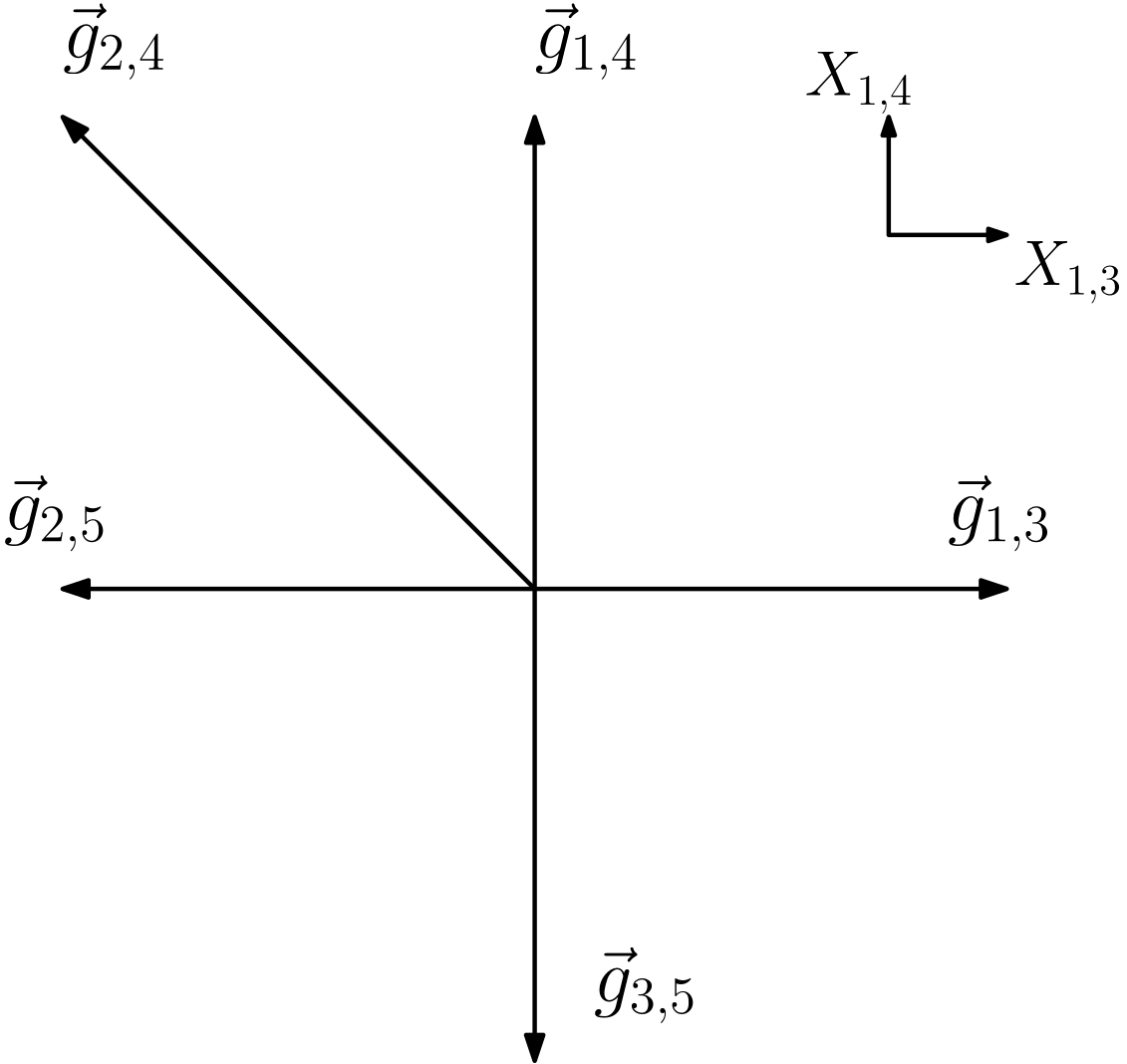}
		\caption{The five-point Feynman fan in the $\{X_{13},X_{14}\}$ basis.}
		\label{fig:5-pt Feynman fan}
	\end{figure}

	The $g$-vector $g_{i,j}$ is seen to be normal to the corresponding facet where $X_{i,j}=0$. The resulting Feynman fan is depicted in figure \ref{fig:5-pt Feynman fan}.
	\be\label{eq:g-vectors 5-pt}\ba \vec{g}_{1,3} = (1,0),\quad \vec{g}_{1,4}=~&(0,1),\quad \vec{g}_{2,4} = (-1,1),\\ \vec{g}_{2,5} = (-1,0)&,\quad \vec{g}_{3,5}=(0,-1).\ea\ee

	When performing a $g$-vector shift, we are effectively moving towards infinity in kinematic space along a specific ray parameterized by $\vec{t}\in\mathbb{R}^{n-3} \setminus \{0\}$. As a result, we will only be able to reach a subset of the boundaries of the associahedron, while others will become inaccessible. In fact, all of the reachable boundaries will have dimension $d>0$, i.e. we won't be able to access any vertices since, by projective invariance, there are no vertices at infinity. The boundaries that we are able to go onto precisely correspond to those partial triangulations containing unshifted poles. 
	
	For example, if we move towards infinity in the $\vec{t}=(0,1)$ direction (i.e. vertically) we will only be able to access the one-dimensional facets associated to $X_{1,3}$ and $X_{2,5}$. In other words, we can only reach boundaries corresponding to partial triangulations of the surface defined by those chords. The shifted amplitude $\ah_n^\infty$ will be a sum over all accessible boundaries after going to infinity.
	
\begin{figure}[ht]
		\centering
		\includegraphics[width=0.23\textwidth,valign=c]{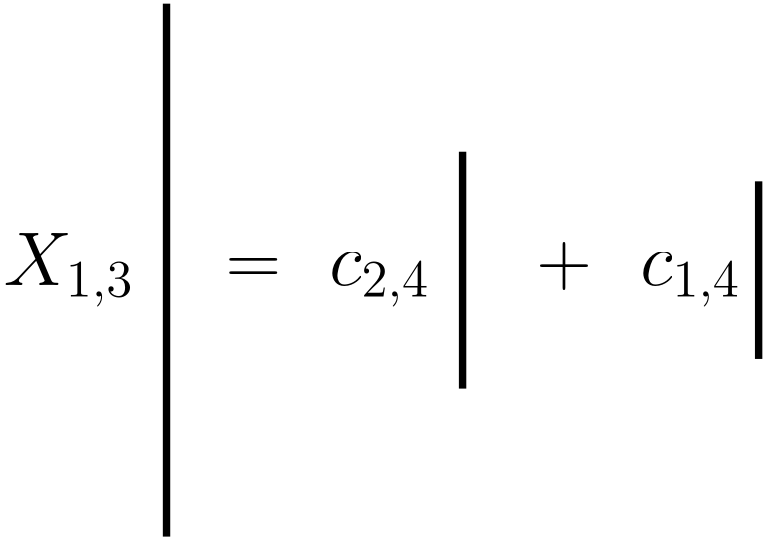}
		\caption{The Minkowski sum decomposition of the $X_{1,3}$ facet; compare to figure \ref{fig:5-pt associahedron}.}
		\label{fig:X13 Minkowski sum}
	\end{figure}

	Moreover, we can associate a numerator factor or contact term to each of these lower codimension boundaries by using their Minkowski sum decompositions. More precisely, the contact term will be given by the total size of the top-dimensional Minkowski summands that constitute the boundary. For instance consider the $X_{1,3}$ facet, which receives contributions from two top-dimensional (in this case, one-dimensional) summands as shown in figure \ref{fig:X13 Minkowski sum}. The contact term for this boundary is therefore $c_{1,4}+c_{2,4}$.
	Similarly, the contact term for the $X_{2,5}$ boundary is $c_{2,4}$. Thus, the shifted amplitude will be given by a sum over those two partial triangulations:
	\be\label{eq: (0,1) shift} \ah_5^{\Phi^3}\xrightarrow[z\to\infty]{\vec{t}=(0,1)} -\frac{1}{z^2}\left( \frac{c_{1,4}+c_{2,4}}{X_{1,3}} + \frac{c_{2,4}}{X_{2,5}} \right).\ee
    This criterion for determining the contact term for a certain boundary is equivalent to the one given in Section \ref{sec:kinematic dependence}.
  Meanwhile, if we moved in a generic direction $\vec{t}=(t_1,t_2)$, we would lose access to all codimension $d>0$ boundaries of the associahedron. Thus, the only contribution will be of the full associahedron itself (corresponding to the empty partial triangulation of the surface). From the Minkowski sum shown in figure \ref{fig:5-pt associahedron}, we see that there is only one full- (in this case, two-dimensional) summand, $c_{1,4}$, so altogether we have
  \be \ah_5^{\Phi^3}\xrightarrow[z\to\infty]{\vec{t}=(t_1,t_2)} -\frac{1}{z^3}
  \frac{1}{t_1(t_1-t_2)t_2} c_{1,4}.\ee

\subsection{\texorpdfstring{$n=6$ associahedron}{n=6 associahedron}}
\begin{figure}[ht]
		\centering
		\includegraphics[width=1\textwidth,valign=c]{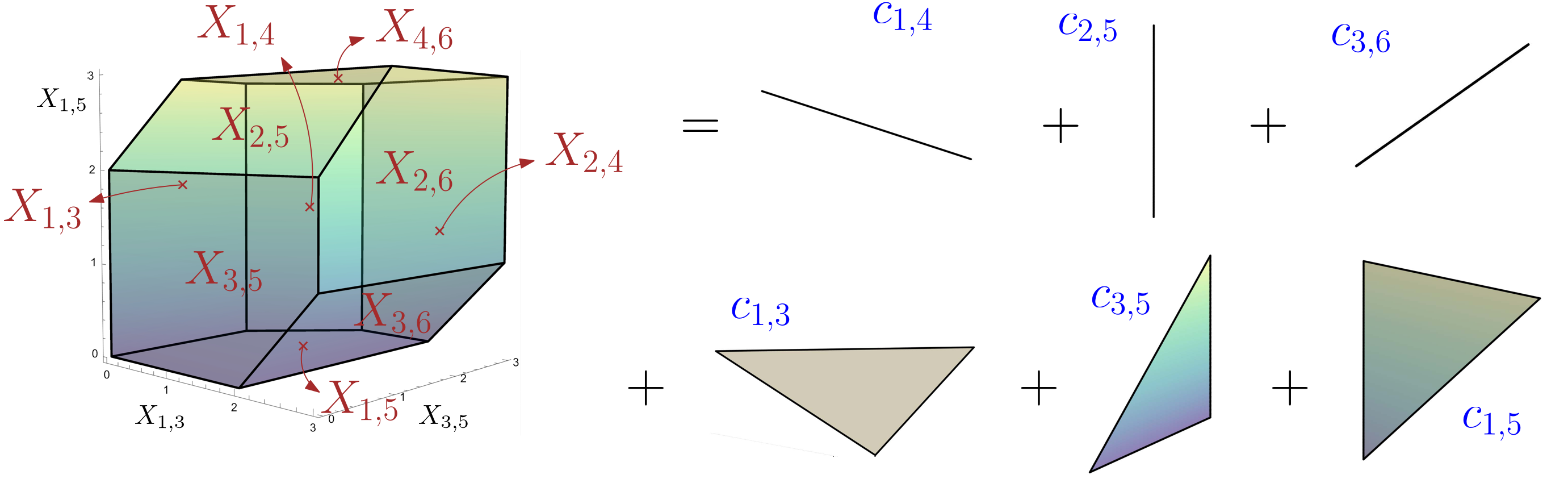}
		\caption{The six-point associahedron in the scaffolding triangulation and its Minkowski summands.}
		\label{fig:6-pt associahedron scaffold}
	\end{figure}

Let's illustrate this procedure via another example. In this section, we understand geometrically the kinematic shift that reproduces the six-point NLSM amplitude \cite{Arkani-Hamed:2023swr}. Consider the kinematic basis given by the ``scaffolding'' triangulation $\{X_{1,3},X_{3,5},X_{1,5}\}$. The remaining non-planar invariants are $\{c_{1,3},c_{1,4},c_{1,5},c_{2,5},c_{3,5},c_{3,6}\}$. The associahedron and its Minkowski sum are depicted in figure \ref{fig:6-pt associahedron scaffold}.
	In this kinematic basis, the rest of the planar variables are given by:
	\be\ba X_{1,4} &= c_{3,5} + c_{3,6}  - X_{3,5}+ X_{1,3},\\ X_{2,4} &= c_{1,3} + c_{3,5} + c_{3,6} - X_{3,5},\\ X_{2,5} &= c_{1,3} + c_{1,4} - X_{1,3} + X_{1,5},\\ X_{2,6} &= c_{1,3} + c_{1,4} + c_{1,5} - X_{1,3},\\ X_{3,6}& = c_{1,5}+c_{2,5}-X_{1,5} + X_{3,5},\\ X_{4,6}&= c_{1,5}+c_{2,5}+c_{3,5} - X_{1,5}.\ea\ee
	Thus, we get the following collection of $g$-vectors:
	\begin{alignat}{3} \vec{g}_{1,3}&=(1,0,0),\quad &\vec{g}_{3,5}&=(0,1,0),\quad &\vec{g}_{1,5}&=(0,0,1),\\ \vec{g}_{1,4}&=(1,-1,0),\quad &\vec{g}_{2,5} &= (-1,0,1),\quad &\vec{g}_{3,6} &= (0,1,-1),\\ \vec{g}_{2,4} &= (0,-1,0),\quad &\vec{g}_{2,6}&=(-1,0,0),\quad &\vec{g}_{4,6}&=(0,0,-1). \end{alignat}
	In order to reproduce the six-point pion amplitude, we want to exclusively preserve the three-particle poles $X_{1,4},X_{2,5},X_{3,6}$. From the expressions for the $g$-vectors, we see that the only shift that accomplishes this is the one along the direction $\vec{t}=(1,1,1)$. If we move towards infinity in that direction, the only facets of the associahedron that will still be reachable are the ones corresponding to the three chords $X_{1,4},X_{2,5},X_{3,6}$. We can then study the Minkowski summands that make up these boundaries. All of them are quadrilaterals, which are themselves spanned by line segments. For example, the facet $X_{1,4}$ is the Minkowski sum of a line segment of length $c_{13}$ and three segments of length $c_{1,5},c_{2,5}$ and $c_{3,5}$ which are orthogonal to the first one (note that some of these summands are higher-dimensional objects in the complete associahedron, but their projection onto the $X_{1,4}$ facet is one-dimensional). It is natural to then associate a numerator $c_{1,3}\times(c_{1,5}+c_{2,5}+c_{3,5})$ to this boundary. A similar analysis for the other two poles yields the following final expression:
	\be\ba \ah_6^{\Phi^3} \xrightarrow[z\to\infty]{\vec{t}=(1,1,1)} \frac{1}{z^4}\Bigg( \frac{c_{1,3}(c_{1,5}+c_{2,5}+c_{3,5})}{X_{1,4}} + \frac{c_{1,5}(c_{3,5}+c_{3,6}+c_{1,3})}{X_{2,5}} \\+ \frac{c_{3,5}(c_{1,3}+c_{1,4}+c_{1,5})}{X_{3,6}} \Bigg) = \ah_6^{\text{NLSM}},\ea\ee
	which is indeed the expression for the six-point pion amplitude in this kinematic basis. Naively, it may seem like we are missing the contact term of the full associahedron/surface, since technically this also contributes at order $\oh(z^{-4})$. However, it turns out that the overall contact term in the scaffolding basis is just zero! Indeed none of the Minkowski summands in figure \ref{fig:6-pt associahedron scaffold} is three-(top-)dimensional, so they don't contribute to the overall contact term. Equivalently, we can see from the six-point mesh in figure \ref{fig:6-pt kinematic mesh} that the maximal causal diamonds with $X_{1,3},X_{3,5}$ and $X_{1,5}$ at their bases don't simultaneously overlap on any square.

\section{Polytope description for mixed amplitudes with two pions}
	\label{sec:polytope}
	
We now turn to make use of the results we have obtained by studying the $g$-vector shifts to provide a geometrical realization of a certain class of mixed amplitudes involving scalars and pions. In particular, we are able to find a family of polytopes which encode the combinatorics of the ordered $n$-point amplitudes involving two pions $\pi$ and $n{-}2$ scalars $\phi$, $\ah_n[\pi_1,\phi_2,\ldots,\phi_{i-1},\pi_i,\phi_{i+1},\ldots,\phi_n]$ for any $i\in \{2,\ldots,n{-}1\}$. In Section  \ref{sec: polytope motivation}, we present general arguments for why such a polytope exists and identify g-vector shifts that can be used to reach them. In Section \ref{sec: polytope diff form}, we formally construct them by characterizing their facets and provide the corresponding differential forms.
	
	\subsection{Combinatoric motivation}\label{sec: polytope motivation}
	
	From what we have observed so far, it is not surprising that these polytopes are simply deformations of some of the boundaries of the full associahedron. Indeed, since these mixed amplitudes arise from performing certain $g$-vector shifts on the $\Tr(\Phi^3)$ scalar amplitude, the corresponding geometry will contain all of the partial triangulations that are still allowed to appear with the set of unshifted chords in the surface.
	
	Note that there are many possible scalar-pion interactions, making it surprising that the one we land on via $g$-vector shifts is the same as that discovered as an extension to the NLSM in the soft limit \cite{Cachazo:2016njl, Mizera:2018jbh}. As explained in \cite{Arkani-Hamed:2024nhp}, the leading contribution in the cubic coupling for these mixed scattering processes involving scalars and pions is given by $\Tr(\Phi^3)$ Feynman diagrams with an even number of $\pi$'s coming off scalar legs. For example, the diagrams contributing to the $\ah_n[\pi_1,\pi_2,\phi_3,\ldots,\phi_n]$ amplitude where the two pions are adjacent are in a one-to-one correspondence with cubic graphs where particles 1 and 2 have merged together into an off-shell leg. In other words, the combinatorics simply corresponds to that of an $(n{-}1)$-point surface, which is realized by removing the poles $X_{1,3}$ and $X_{2,i}$ ($i=4,\ldots,n$). From the kinematic mesh, it is clear to see that such an outcome can be achieved by considering the kinematic basis associated to the ray-like triangulation $\{X_{1,3},X_{1,4},\ldots,X_{1,n-1}\}$ and performing a $g$-vector shift in the direction $\vec{t}=(1,0,\ldots,0)$.
	
	As a result, we conclude that the polytope describing the full kinematic dependence of the mixed amplitude with two adjacent pions is an $(n{-}1)$-point associahedron, where each vertex has been dressed by a numerator factor. These numerators are given by the contact terms of the corresponding partial triangulations, as seen in Section \ref{sec:kinematic dependence}. This is also evident by starting with the full $\Tr(\Phi^3)$ associahedron in the ray-like basis. Indeed, the geometry encoding the combinatorics of the partial triangulations is given by the projection of the reachable facets onto the $X_{1,3}$ direction. This is by construction isomorphic to the $X_{1,3}$ facet itself, i.e. an $n{-}1$-point associahedron.
	
\begin{figure}[ht]
		\centering
		\includegraphics[width=0.3\textwidth,valign=c]{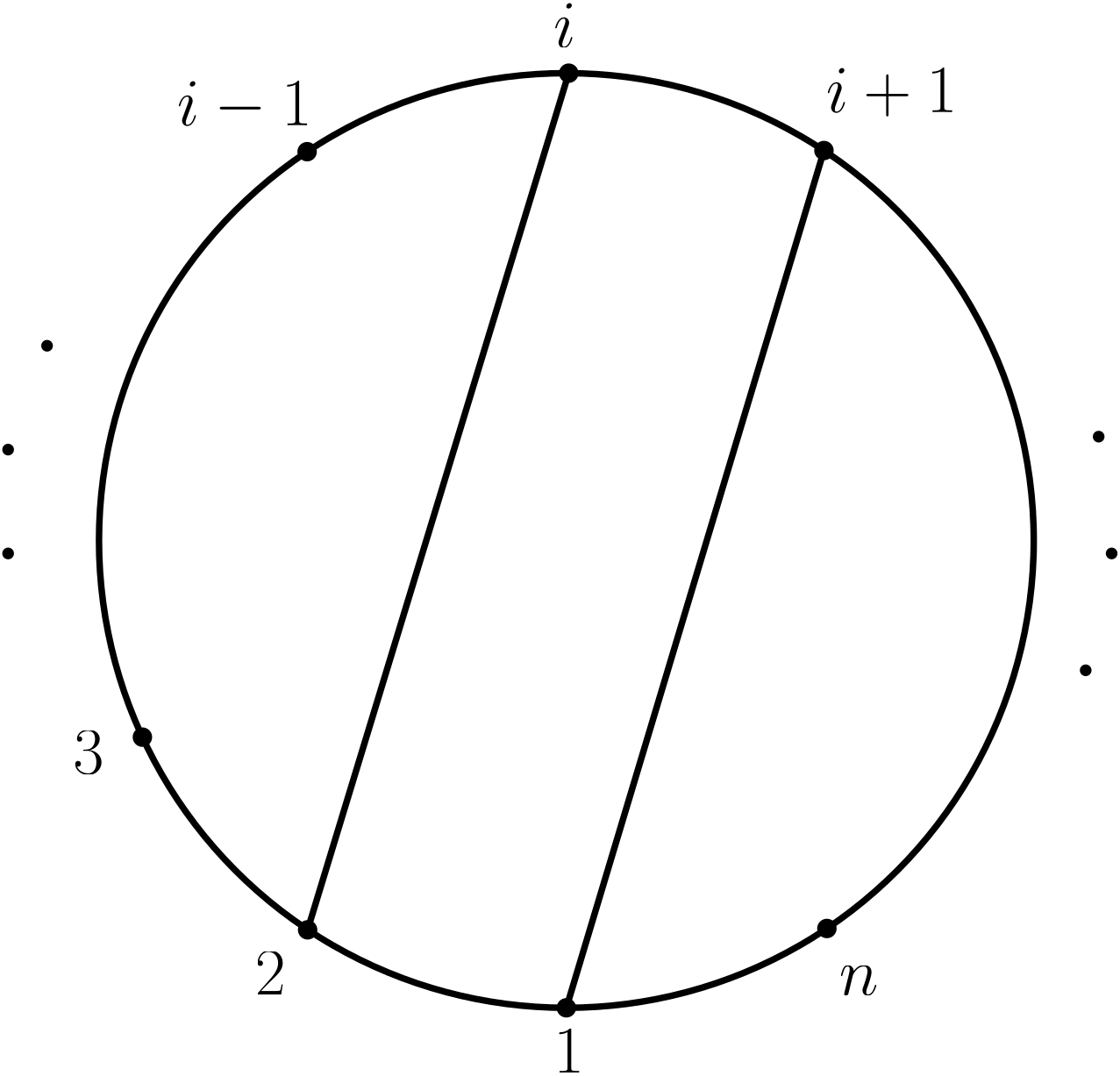}
		\caption{Effective subsurfaces for the $\ah_n[\pi_1,\phi_2,\ldots,\pi_i,\ldots,\phi_n]$ mixed amplitude.}
		\label{fig:subsurfaces non-adjacent pions}
	\end{figure}

	For any other case $\ah_n[\pi_1,\phi_2,\ldots,\pi_i,\ldots,\phi_n]$ in which the pions are non-adjacent, the diagrams contributing to leading order in the coupling will always contain the propagators $X_{1,i+1}$ and $X_{2,i}$. This leaves us with the scalar combinatorics of two subsurfaces $(2,3,\cdots, i)$ and $(1,\,i{+}1,\cdots, n)$ (see figure \ref{fig:subsurfaces non-adjacent pions}).
	Using the same arguments as before, the polytope that determines the full kinematic dependence of this amplitude is the direct product of the corresponding lower-point associahedra, where the vertices are again dressed with the appropriate numerator factors.
	
	As for the shifts that generate these amplitudes, we found the following non-unique prescription: the base triangulation contains all the chords $X_{3,j}$ with $j=i{+}2,\ldots,n,1$ and all $X_{k,i+2}$ with $k=4,\ldots,i{+}1$. The $g$-vector shift given by $\vec{t}=(1,1,\ldots,1)$ then generates the mixed amplitude $\ah_n[\pi_1,\phi_2,\ldots,\pi_i,\ldots,\phi_n]$.
	
	The fact that the vertices of the polytopes describing this family of mixed amplitudes have to be dressed with non-trivial numerator factors implies that the corresponding differential form is no longer guaranteed to be projectively invariant (in fact, in some cases it is not). As a result, there will be real poles at infinity that don't cancel out in the whole amplitude, and which are not spuriously introduced by the Feynman diagram expansion. However, this shouldn't really come as a surprise when we think of the origin of these amplitudes from the perspective of the kinematic shifts, since they precisely tell us that these objects are remnants of the $\Tr(\Phi^3)$ amplitude \emph{at infinity}!
	
	\subsection{Construction of the differential form}\label{sec: polytope diff form}
    
	The facet description of the polytopes for the mixed amplitudes involving two pions is straightforward: take all the forbidden chords that are shifted away to infinity and set them to positive constants:
	\be X_{i,j}^\text{forbidden} := b_{i,j}>0.\ee
As mentioned in the previous paragraph, the combinatorics of these amplitudes are described by a set of lower-point surfaces (where the specifics depend on the configuration of the pions in the color ordering). For the rest of the planar invariants living inside these lower-point surfaces, we can simply impose the usual positivity conditions of the associahedron as described in Section \ref{sec: review kin space}. In other words, the polytope will be carved out by the intersection of the simplex $X_{i,j}^{\rm allowed}>0$ and the subspaces $H_{k}^i$ for each lower-point surface, which are defined by choosing a kinematic basis on them and setting the non-planar invariants to positive constants, $c_{k,l}^{S_i}>0$.
	
    Meanwhile, the expression for the mixed amplitude is encoded in the usual way that a differential form $\Omega(\mathcal{P})$ is associated to any positive geometry:
	\be \Omega(\mathcal{P}) = \left(\frac{1}{z^2}\ah_n^\infty\right)\, d^{n-4}X.\ee
	The polytopes describing the mixed amplitudes are simple, since the partial triangulations of the surface corresponding to the Feynman diagrams always have the same number of chords, $n{-}4$. Thus, the different vertices are adjacent to exactly $n{-}4$ facets. As explained in \cite{Arkani-Hamed:2017mur}, this means that the differential form can be expressed as a sum over the vertices $Z$ of the polytope:
	\be\label{eq:differential form} \Omega = \sum_Z \text{sgn}(Z) \nh(Z) \bigwedge_{a=1}^{n-4} d\log X_{i_a,j_a},\ee
	where $(1,X)\in\mathbb{P}^{n-4}$, $\nh(Z)$ is the numerator factor associated to each vertex and $X_{i_a,j_a}$ are the chords appearing in the corresponding partial triangulation of the surface. We emphasize that due to these numerator factors, the differential forms associated to our mixed amplitude polytopes are not independent of the choice of triangulation and are therefore not canonical forms.
	
\begin{figure}[ht]
		\centering
		\includegraphics[width=0.38\textwidth,valign=c]{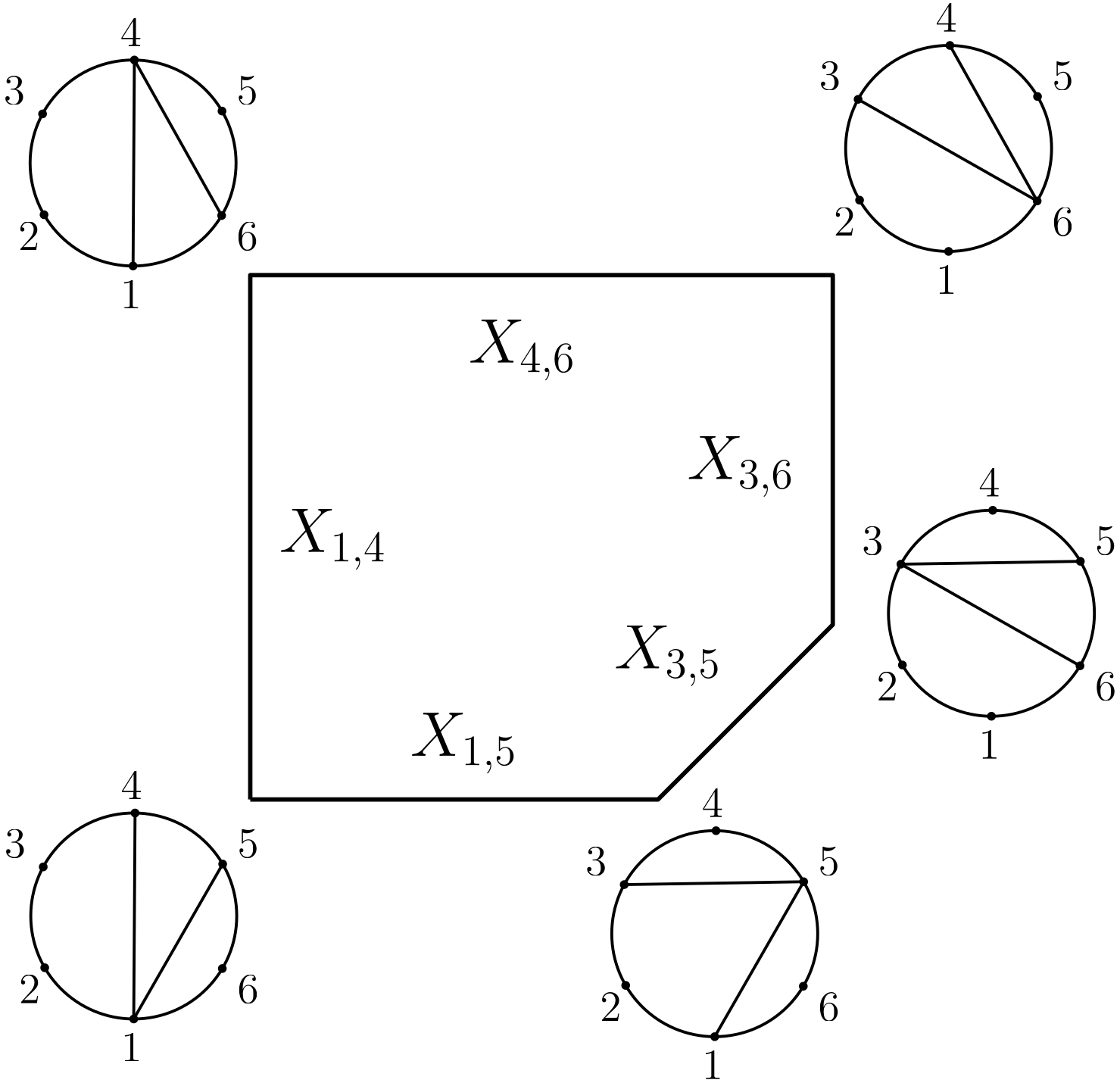}
		\caption{The polytope for the mixed amplitude $\ah_n[\pi_1,\pi_2,\phi_3,\phi_4,\phi_5]$, equivalent to a five-point associahedron. Each vertex corresponds to a partial triangulation made of unshifted chords.}
		\label{fig:mixed amp pion12}
	\end{figure}

	Let's illustrate this with an example: consider the six-point mixed amplitude with two adjacent pions $\ah_6[\pi_1,\pi_2,\phi_3,\phi_4,\phi_5,\phi_6]$. The forbidden chords are $X_{1,3},X_{2,4},X_{2,5}$ and $X_{2,6}$. In our polytope, they are set to constants:
	\be X_{1,3}=b_{1,3}>0,\ X_{2,4}=b_{2,4}>0,\ X_{2,5}=b_{2,5}>0,\ X_{2,6}=b_{2,6}>0.\ee
	Meanwhile, for the effective subsurface ((1,2),3,4,5,6), we impose the usual conditions of the associahedron:
	\be \ba X_{1,4},X_{1,5},X_{3,5}&,X_{3,6},X_{4,6}>0,\\ c_{1,4}+c_{2,4} = X_{1,4}& - X_{1,5} + X_{3,5} >0,\\ c_{1,5}+c_{2,5} = X_{1,5}&-X_{3,5}+X_{3,6}>0,\\ c_{3,5}=X_{3,5}&-X_{3,6}+X_{4,6}>0.\ea\ee
	Pulling back onto the $(X_{1,4},X_{1,5})$ space, these conditions become:
	\be \ba  0<X_{1,4}<c_{1,4}+c_{2,4}+c_{1,5}+c_{2,5},\\ 0<X_{1,5}<c_{1,5}+c_{2,5}+c_{3,5},\\ X_{1,4}-X_{1,5}<c_{1,4}+c_{2,4},\ea\ee
	which indeed carves out a two-dimensional associahedron (see figure \ref{fig:mixed amp pion12}).
	Using the expression for the differential form associated to this polytope as a sum over its vertices \eqref{eq:differential form}, we indeed recover the correct expression for the amplitude:
	\begin{multline} \ah_6[\pi_1,\pi_2,\phi_3,\phi_4,\phi_5,\phi_6]=\\  \frac{c_{1,3}}{X_{1,4}X_{1,5}} + \frac{c_{1,3}+c_{1,4}}{X_{1,5}X_{3,5}} + \frac{c_{1,3}+c_{1,4}+c_{1,5}}{X_{3,5}X_{3,6}} + \frac{c_{1,3}+c_{1,4}+c_{1,5}}{X_{3,6}X_{4,6}}  + \frac{c_{1,3}}{X_{4,6}X_{1,4}}.\end{multline}
	
\section{Discussion}
\label{sec:discussion}
	
In this work, we characterized the functions that result from large kinematic deformations of Tr$(\Phi^3)$ amplitudes under what we call $g$-vector shifts. These functions at infinity have many interesting properties: they consistently factorize onto themselves, display enhanced large $z$ fall off, and generic $g$-vector shifts commute with one another. A geometric construction of these functions at infinity was also presented. Large deformations in special directions lead to amplitudes in a theory of scalars coupled to pions. These amplitudes can be described geometrically via a deformed $n$-point associahedron that shares the same combinatorial structure as a single, or a product of two, lower-point associahedra. 

Indeed our main result \eqref{eq:mainresult} displays another interesting feature: given a kinematic basis, there exists an expression for the large deformation where only $c$ variables in the basis appear in the numerator. This is trivial in the case of the un-deformed Tr$(\Phi^3)$ theory (which has unit numerator), but non-trivial in its deformations that have numerators of higher mass dimension. The existence of such a ``$c$-expansion'' for NLSM for example, is intriguing and motivates the question of what such a property means for theories defined on the surface and how universal it is.

It is important to note that though the discussion in this work is limited to tree-level amplitudes in Tr$(\Phi^3)$ theory, $g$-vector shifts can be straightforwardly generalized to higher loop orders and to other theories defined on kinematic surfaces, such as YM theory \cite{Arkani-Hamed:2023jry,Arkani-Hamed:2024tzl}. If the numerator structure can be determined in a similar way as we have for trees in Tr$(\Phi^3)$ theory, then there are many interesting directions for future study, in particular the behavior of the stringy deformation of Tr($\Phi^3$), discussed in \cite{Arkani-Hamed:2023swr,Cao:2024gln,Cao:2024qpp,Dong:2024klq} and the analogous geometric interpretation of the large deformations of the 1-loop surfacehedron \cite{Arkani-Hamed:2023mvg, Arkani-Hamed:2023lbd}.

\begin{figure}[ht]
	\centering
	\includegraphics[width=0.65\textwidth,valign=c]{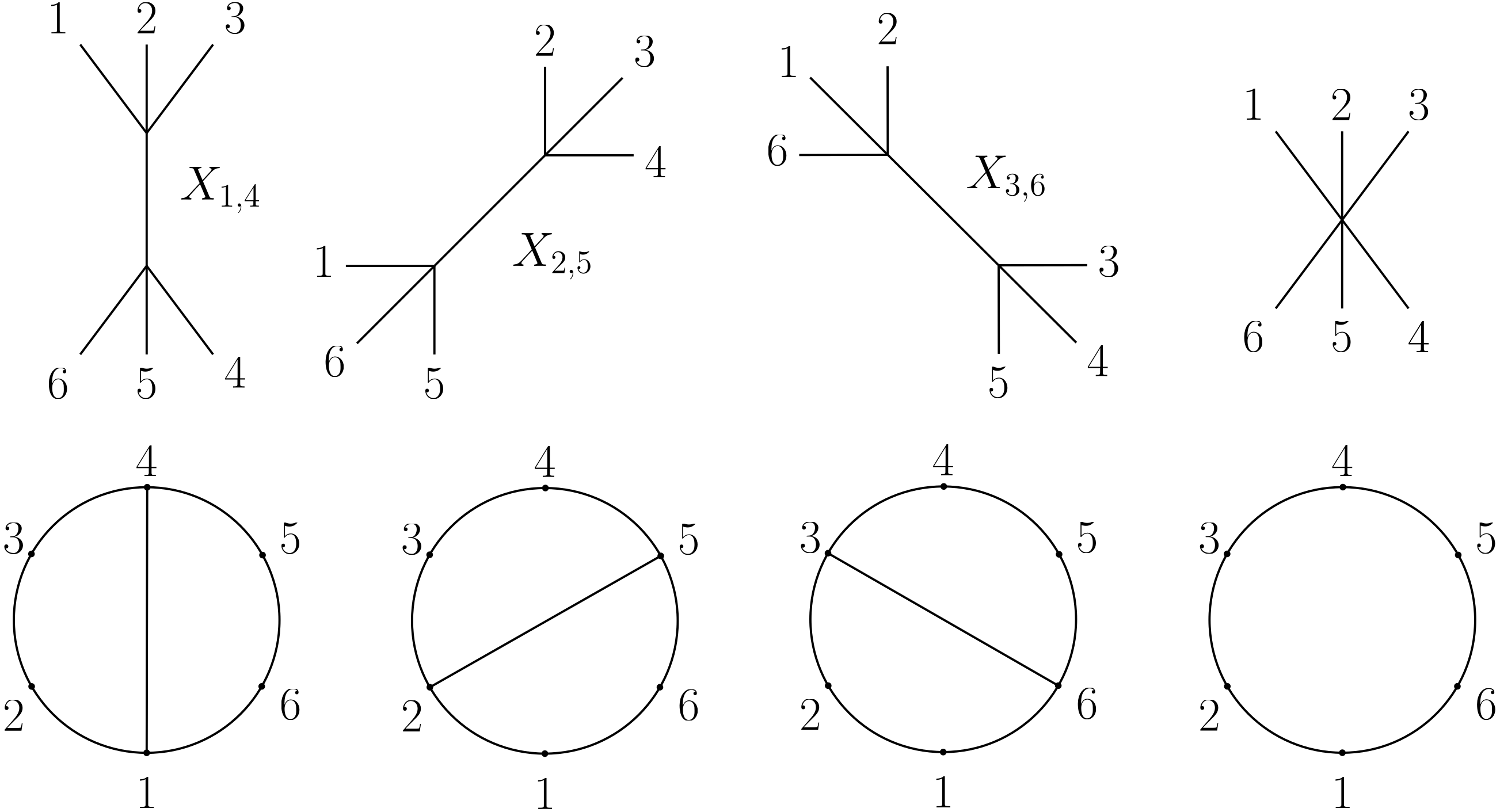}
	\caption{Feynman diagrams contributing to the 6-point NLSM amplitude, together with the corresponding partial triangulations of the surface.}
	\label{fig:6-pt NLSM graphs}
\end{figure}

Finally, we discuss the polytopes introduced in Section \ref{sec:polytope}. These geometries describe mixed amplitudes involving two pions and any number of scalars, and it is natural to wonder whether we could perform a similar construction for mixed amplitudes with any number of pions. However, there seems to be an obstruction to finding polytopal realizations of the combinatorics involving a higher number of pions. To see this, we simply need to look at the six-point NLSM amplitude $\ah_6^{\text{NLSM}}$. Since the model includes all even-point interactions, in an arbitrary kinematic basis there are four Feynman diagrams contributing to this amplitude, as depicted in figure \ref{fig:6-pt NLSM graphs}.
Considering there is at most one propagator in these diagrams, we expect that a geometry describing this process would be one-dimensional. However, there is clearly no one-dimensional closed geometry with four vertices. 

At higher points, these issues start to accumulate, since one needs to take into account all possible lower-point contact terms corresponding to different even-point interactions. Already in the case with four pions, there are subsets of Feynman diagrams reproducing the combinatorics of the six-point NLSM amplitude. Thus finding a polytope description of the former requires also understanding the geometric realization of the latter. While there have been attempts to find polytopal realizations of other theories with scalar interactions \cite{Banerjee:2018tun,Raman:2019utu,Herderschee:2019wtl,Aneesh:2019cvt,Aneesh:2019ddi,Herderschee:2020lgb,Jagadale:2021iab}, the NLSM still evades description because of the reasons discussed above, and we leave the resolution for future work.

	\section*{Acknowledgements}
	
	We thank N.~Arkani-Hamed for suggesting that a world at infinity might exist, and we are grateful to C.~Figueiredo and A.~Laddha for many helpful discussions. This work was supported in part by the US Department of Energy under contract DE-SC0010010 Task F (MSp, AV) and by Simons Investigator Award \#376208 (SP, AV).

\bibliographystyle{JHEP}
	

\providecommand{\href}[2]{#2}\begingroup\raggedright\endgroup

\end{document}